\documentclass{bioinfo}
\copyrightyear{2017} \pubyear{2017}

\access{Advance Access Publication Date: Day Month Year}
\appnotes{Review Article}

\begin{document}
\firstpage{1}

\subtitle{Subject Section}

\title[Cancer Driver Discovery]{Computational methods for cancer driver discovery: A survey}
\author[Pham et al.]{Vu Viet Hoang Pham\,$^{\text{\sfb 1}}$, Lin Liu\,$^{\text{\sfb 1}}$, Cameron Bracken\,$^{\text{\sfb 2, 3}}$, Gregory Goodall\,$^{\text{\sfb 2, 3}}$, Jiuyong Li\,$^{\text{\sfb 1}}$, and Thuc Duy Le\,$^{\text{\sfb 1,}*}$}
\address{$^{\text{\sf 1}}$UniSA STEM, University of South Australia, Mawson Lakes, SA 5095, AU, \\
$^{\text{\sf 2}}$Centre for Cancer Biology, SA Pathology, Adelaide, SA 5000, AU, and \\
$^{\text{\sf 3}}$Department of Medicine, The University of Adelaide, Adelaide, SA 5005, AU.}

\corresp{$^\ast$To whom correspondence should be addressed.}

\history{Received on XXXXX; revised on XXXXX; accepted on XXXXX}

\editor{Associate Editor: XXXXXXX}

\abstract{\textbf{Motivation:} Uncovering the genomic causes of cancer, known as cancer driver genes, is a fundamental task in biomedical research. Cancer driver genes drive the development and progression of cancer, thus identifying cancer driver genes and their regulatory mechanism is crucial to the design of cancer treatment and intervention. Many computational methods, which take the advantages of computer science and data science, have been developed to utilise multiple types of genomic data to reveal cancer drivers and their regulatory mechanism behind cancer development and progression. Due to the complexity of the mechanistic insight of cancer genes in driving cancer and the fast development of the field, it is necessary to have a comprehensive review about the current computational methods for discovering different types of cancer drivers.\\
\textbf{Results:} We survey computational methods for identifying cancer drivers from genomic data. We categorise the methods into three groups, methods for single driver identification, methods for driver module identification, and methods for identifying personalised cancer drivers. We also conduct a case study to compare the performance of the current methods. We further analyse the advantages and limitations of the current methods, and discuss the challenges and future directions of the topic. In addition, we investigate the resources for discovering and validating cancer drivers in order to provide a one-stop reference of the tools to facilitate cancer driver discovery. The ultimate goal of the paper is to help those interested in the topic to establish a solid background to carry out further research in the field.
\textbf{Contact:} \href{Thuc.Le@unisa.edu.au}{Thuc.Le@unisa.edu.au}\\
}

\maketitle


\section{Introduction}



Identifying cancer driver genes (cancer drivers for short) is vital since these genes play a significant role in the development of cancer. Understanding cancer drivers and their regulatory mechanism is crucial to the design of effective cancer treatments.

\begin{figure}[h!]
\centering
\includegraphics[width=0.3\textwidth]{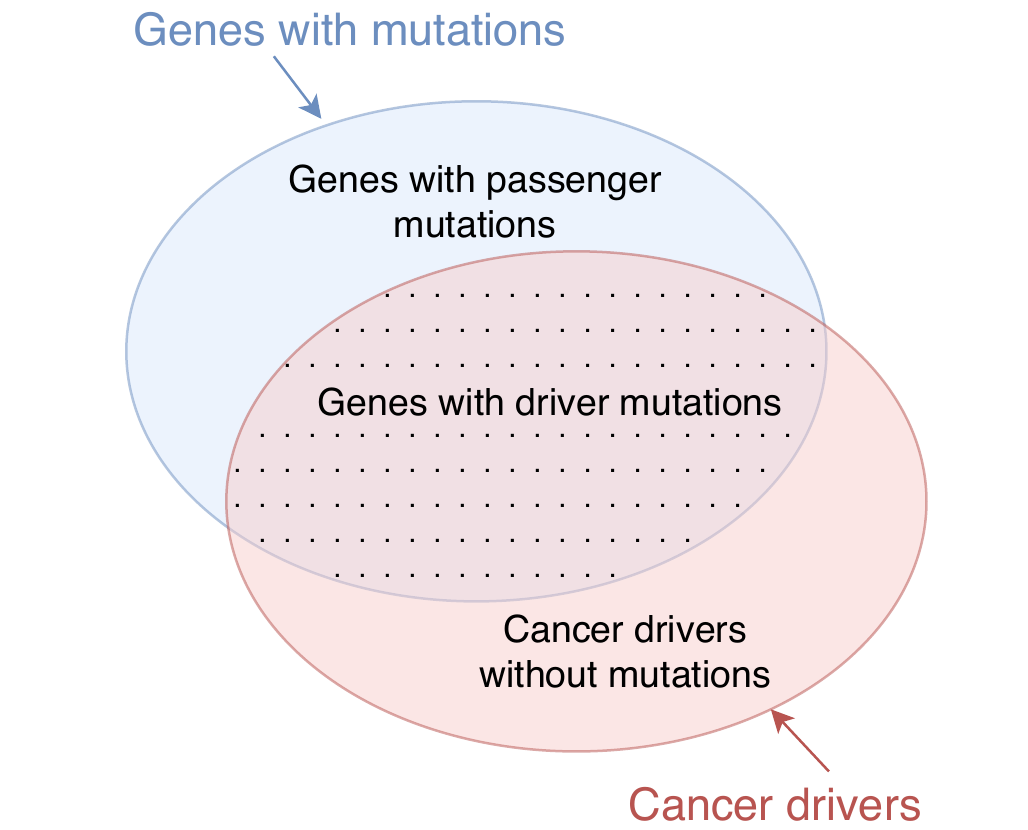}
\caption{Cancer drivers and genes with mutations. Genes with driver mutations are cancer drivers. Some genes which do not contain mutations but regulate driver mutations to develop cancer are also considered as cancer drivers.}
\label{fig:CancerDriver}
\end{figure}

Classical methods of identifying cancer driver genes are based on detecting the mutations in the DNA sequences of coding genes in wet-lab experiments. There are many mutation types in the genome such as single-nucleotide variants (SNVs), structural variants (SVs), insertions and deletions (indels), and copy number aberrations (CNAs) \citep{XRN239}. These mutations may cause normal cells to transform to tumour cells, resulting in the development of cancer. For example, it has been confirmed that mutations in genes \textit{VHL} and \textit{MET} cause kidney cancer \citep{XRN247} and mutations in genes \textit{AKT1} and \textit{BRCA1} are related to breast cancer \citep{XRN248}. However, many mutated genes are not driver genes and may not regulate the progression of cancer. The reason is that not all mutations in the genome contribute to cancer development. Mutations which play a significant role in cancer progression are called driver mutations while mutations which do not have any impact on cancer development are called passenger mutations \citep{XRN235, XRN233}. Genes which bear cancer driver mutations are considered as cancer drivers \citep{Tokheim14330}. Nevertheless, some cancer drivers may not contain mutations. For example, genes which may not contain mutations but regulate targets to develop cancer are also considered as cancer driver, e.g. the overexpression of \textit{KDM5C} decreases p54 expression to enhance the proliferation and invasion of gastric cancer cells and \textit{KDM5C} is considered as a cancer driver \citep{HRN16}. The illustration of cancer drivers and genes with mutations is shown in Figure \ref{fig:CancerDriver}.

Given the complexity of the regulation by cancer drivers and the large number of genes, over ten thousand, detecting cancer driver genes is challenging with the wet-lab experiments and many computational methods utilising multiple types of genomic data have been developed to reveal cancer drivers and their regulatory mechanism behind the cancer development \citep{XRN192, XRN177, XRN230, XRN162}. Cancer driver discovery methods are increasingly popular recently because of the fast development of computer science and significant revolution of DNA sequencing techniques. Taking these advantages, numerous methods have been proposed to detect cancer driver genes. For example, MutSigCV \citep{RN233} investigates the significance of mutations in genes to predict cancer drivers, OncodriveFM \citep{XRN245} and OncodriveCLUST \citep{XRN246} evaluate the functional influence and clustering of gene mutations respectively, DriverNet \citep{RN241}, MEMo \citep{XRN252}, and CBNA \citep{RN228} examine the role of genes in gene regulatory networks. Due to the large number of the current computational methods for cancer driver discovery, it may take the huge amount of effort for people to find a good resource to know the state-of-the-art methods, and thus a review is necessary and helpful.



There have been previous works \citep{XRN239} reviewing the computational methods for identifying single cancer drivers at the population level. However, it is important to gain mechanistic insight into how cancer drivers work together in driving cancer. Besides, cancer drivers of each patient may be different from others since cancer is a heterogeneous disease, each patient has a different genome and the disease of each patient may be driven by different cancer driver genes. Thus, we also need to consider cancer driver modules and personalised cancer drivers (i.e. cancer drivers for a specific patient). In addition, there are numerous new cancer driver identification methods which have been developed since then. Therefore, it is required to have a more comprehensive review about the current computational methods for identifying cancer drivers.

In this paper, we survey computational methods for discovering both single cancer drivers and cancer driver modules at the population level and the individual level as well. We then analyse the advantages/disadvantages of the current methods and identify challenges of the field. To facilitate the development of new computational methods for cancer driver detection, we survey resources which can be used as tools in conducting cancer driver research and validating predicted cancer drivers. In addition, with the case study conducted to compare the performance of the current methods in this paper, we believe it will be useful for researchers, who are interested or work in the field, to develop their new methods.

The paper is structured as follows. In Section \ref{section:methods}, we review computational methods for identifying single and cancer driver modules from genomic data, including cancer drivers for both the population and individuals. We summarise the current available sources which can be used for conducting cancer driver researches as well as validating the results in Section \ref{section:resources}. In Section \ref{section:casestudy}, we carry out a case study. Finally, we analyse the current methods to identify their advantages and limitations then discuss future directions and challenges of the field in Section \ref{section:gap}.


\section{Cancer driver discovery methods} \label{section:methods}

\begin{figure}[h!]
\centering
\includegraphics[width=0.5\textwidth]{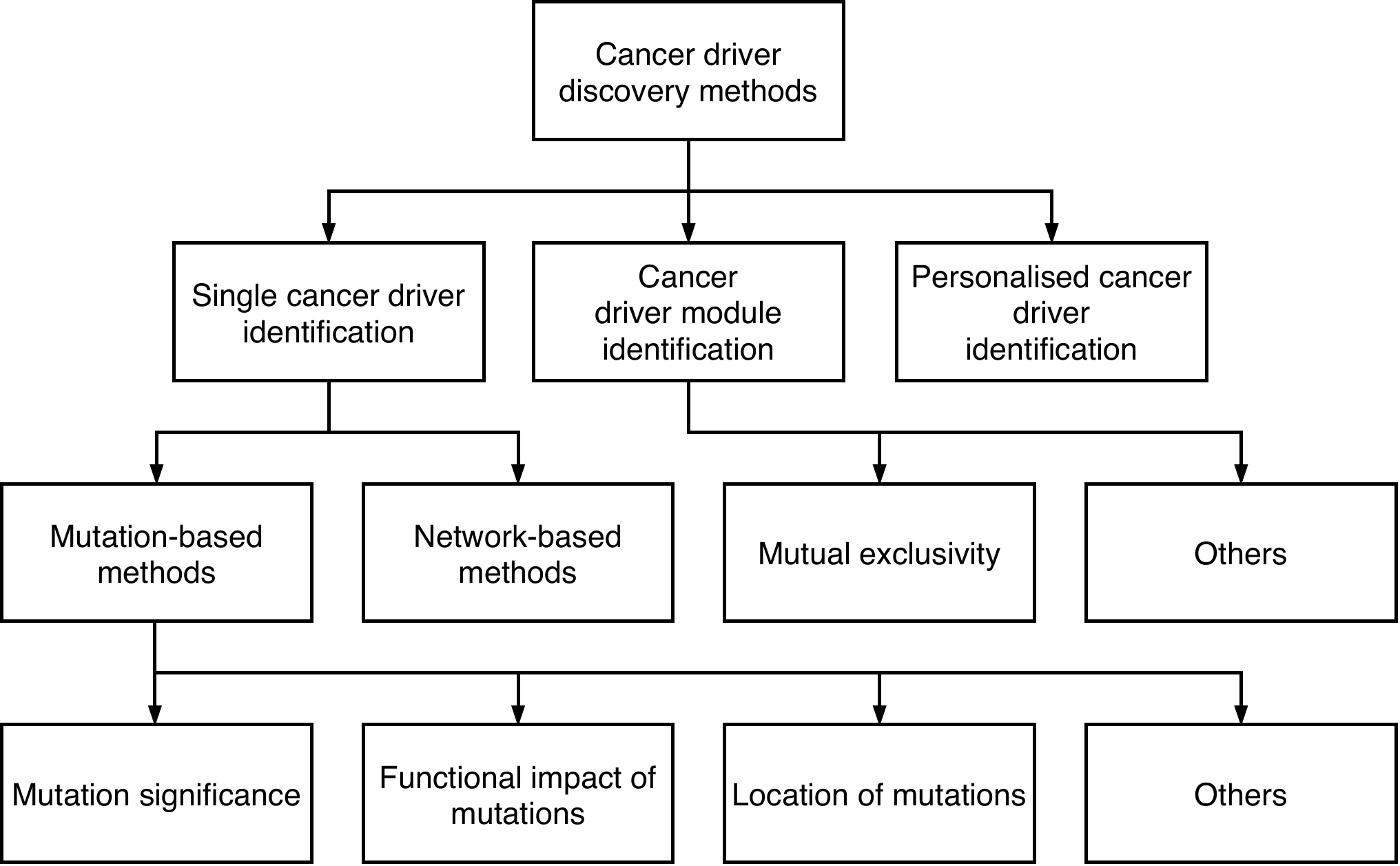}
\caption{Categorisation of cancer driver discovery methods. The methods are categorised in three groups: Single cancer driver identification, Cancer driver module identification, and Personalised cancer driver identification. Single cancer driver identification includes two sub-groups: Mutation-based methods and Network-based methods. Mutation-based methods discover cancer drivers using mutation significance, functional impact of mutations, location of mutations, etc. Most cancer driver module identification methods use the mutual exclusivity of mutations to identify modules of cancer drivers.}
\label{fig:chart}
\end{figure}

The current computational methods use a wide range of genomic data types, including mutations, gene expression, pathways, etc. to discover different types of cancer drivers. Thus, we categorise the methods into various categories and sub-categories. The diagram of the categorisation is shown in Figure \ref{fig:chart} and the summary of the methods is presented in Table~\ref{Tab:02}.

\newcommand{\specialcell}[2][c]{%
\begin{tabular}[#1]{@{}l@{}}#2\end{tabular}}

\begin{table*}[!ht]
\processtable{Summary of methods for identifying cancer drivers \label{Tab:02}} {\begin{tabular}{llll}\toprule 
Category&Sub-category&Method& Description\\\midrule
\specialcell[t]{Single cancer driver\\identification} 
&\specialcell[t]{Mutation-based methods\\(Using mutation significance)}& MutSigCV \citep{RN233} & \specialcell[t]{Assesses the significance of mutations in DNA sequencing in order to discover\\cancer driver genes.}\\
&\specialcell[t]{Mutation-based methods\\(Using functional impact\\of mutations)}& OncodriveFM \citep{XRN245}& \specialcell[t]{Uses the functional impact of mutations of genes to detect cancer drivers with the\\ hypothesis that any bias of variations with a significantly functional impact in genes\\ can be used to identify candidate driver genes.}\\
&& OncodriveFML \citep{XRN344} & \specialcell[t]{Uses the functional impact of gene mutations to reveal both coding and non-coding\\cancer drivers.}\\
&& DriverML \citep{RN232} & \specialcell[t]{Uses the functional impact of mutations to unravel cancer drivers through a\\supervised machine learning approach.}\\
&\specialcell[t]{Mutation-based methods\\(Using location of mutations)}& ActiveDriver \citep{RN234} & \specialcell[t]{Looks at the enrichment of mutations in externally defined regions to uncover cancer\\driver genes.}\\
&& OncodriveCLUST \citep{XRN246} & \specialcell[t]{Detects cancer genes with a large bias in clustering mutations based on the idea\\ that gain-of-function mutations usually cluster in particular protein sections and these\\ mutations contribute to the development of cancer cells.}\\
&\specialcell[t]{Mutation-based methods\\(Others: Combining with\\gene expression, pathways)}& IntOGen-mutations \citep{XRN234} & \specialcell[t]{Uses somatic mutations, gene expression, and tumour pathways to identify cancer\\ drivers for various tumour types by combining OncodriveFM \citep{XRN245} and\\OncodriveCLUST \citep{XRN246}.}\\
&& PathScan \citep{XRN253} & \specialcell[t]{Combines genomic mutations with the information of genes in known pathways to\\uncover cancer driver genes.}\\
&& Sakoparnig et al. \citep{XRN257} & \specialcell[t]{Introduces a computational method to detect genomic alterations with low\\ occurrence frequencies based on mutation timing.}\\
&& CONEXIC \citep{XRN226} & \specialcell[t]{Applies a score-guided search to detect combinations of modulators which reflect\\ the expression of a gene module in a set of tumour samples then it identifies those\\ which have the highest score in amplified or deleted regions.}\\
&& ncDriver \citep{XRN238} & \specialcell[t]{Screens non-coding mutations with conservations and cancer specificity to reveal\\non-coding cancer drivers.}\\
&\specialcell[t]{Network-based methods}& Vinayagam et al. \citep{XRN165} & \specialcell[t]{Applies controllability analysis on the directed network of human protein-protein\\ interaction to identify disease genes.}\\
&& CBNA \citep{RN228} & \specialcell[t]{Identifies coding and miRNA cancer drivers by analysing the controllability of the\\miRNA-TF-mRNA network and mutation data.}\\
&& DriverNet \citep{RN241} & \specialcell[t]{Uncovers cancer drivers by evaluating the influence of mutations on transcriptional\\networks in cancer.}\\
\specialcell[t]{Cancer driver module\\identification}
&\specialcell[t]{Using mutual exclusivity\\of mutations}& CoMEt \citep{XRN235} & \specialcell[t]{Identifies cancer genes by using the exact statistical test to test mutual exclusivity\\ of genomic events and applies techniques to do simultaneous analysis for mutually\\ exclusive alterations.}\\
&& WeSME \citep{RN242} & \specialcell[t]{Discovers cancer drivers by evaluating the mutual exclusivity of mutations of gene\\pairs.}\\
&& MEMo \citep{XRN252} & \specialcell[t]{Analyses mutual exclusivity of mutated genes in subnetworks to identify mutual\\exclusivity modules in cancer.}\\
&\specialcell[t]{Others: Using mutations,\\gene expression, gene network}& iMCMC \citep{XRN228} & \specialcell[t]{Uses the cancer genomic data including mutations, CNAs, and gene expression from\\cancer patients to identify mutated core modules in cancer.}\\
&& NetBox \citep{XRN255} & \specialcell[t]{Uses biological networks to assess network modules statistically and identify core\\pathways in GBM.}\\
&& TieDIE \citep{XRN254} & \specialcell[t]{Applies network diffusion to discover the relationship of genomic events and changes\\in cancer subtypes.}\\
&& Hamilton et al. \citep{XRN236} & \specialcell[t]{Uses the pan-cancer dataset of TCGA and the miRNA target data of AGO-CLIP to\\detect a pan-cancer oncogenic miRNA superfamily with a central core seed motif.}\\
\specialcell[t]{Personalised cancer driver\\identification}
&& DawnRank \citep{XRN237} & \specialcell[t]{A ranking framework which applies PageRank to evaluate the impact of genes in\\ an interaction network to detect cancer drivers.}\\
&& SCS \citep{RN243} & \specialcell[t]{Detects the minimal set of mutated genes controlling the maximal differentially\\expressed genes as cancer drivers.}\\
&& PNC \citep{RN244} & \specialcell[t]{Identifies cancer drivers as the minimum gene set which covers all the edges based on\\a bipartite graph.}\\
\botrule
\end{tabular}}{}
\end{table*}

In the categorisation, we differentiate single cancer drivers from modules of cancer drivers since there is evidence showing that some genes work in concert to influence different biological processes (e.g. EMT) \citep{XRN326} and in some biological processes, the regulation of single genes might not have significant impacts but the regulation of groups of genes does. Furthermore, as cancer is a heterogeneous disease, each patient may have a different morphology and clinical outcome. For instance, two patients, who have the same cancer type and receive the same treatment, may experience different outcomes. The reason is that the genome of each patient is different and each patient's disease may be driven by different driver genes, leading to a strong need to study cancer driver genes specific to an individual patient. Thus, we categorise the current computational methods for cancer driver discovery into three groups, including methods to identify single cancer drivers, methods to identify cancer driver modules, and methods to discover personalised cancer drivers (i.e. cancer drivers for a specific patient). In addition, based on the key techniques used in the methods, we divide single cancer driver identification methods into two sub-groups, including mutation-based methods and network-based methods. Mutation-based methods use different characteristics of mutations such as mutation significance, functional impact of mutations, location of mutations to discover cancer drivers while network-based methods evaluate the role of genes in biological networks to predict cancer drivers. Most of cancer driver module identification methods use the mutual exclusivity of mutations to identify modules of cancer drivers. We will discuss the detail of the methods in the following sections.

\subsection{Single cancer driver identification} \label{single}

\begin{figure}[h!]
\centering
\includegraphics[width=0.3\textwidth]{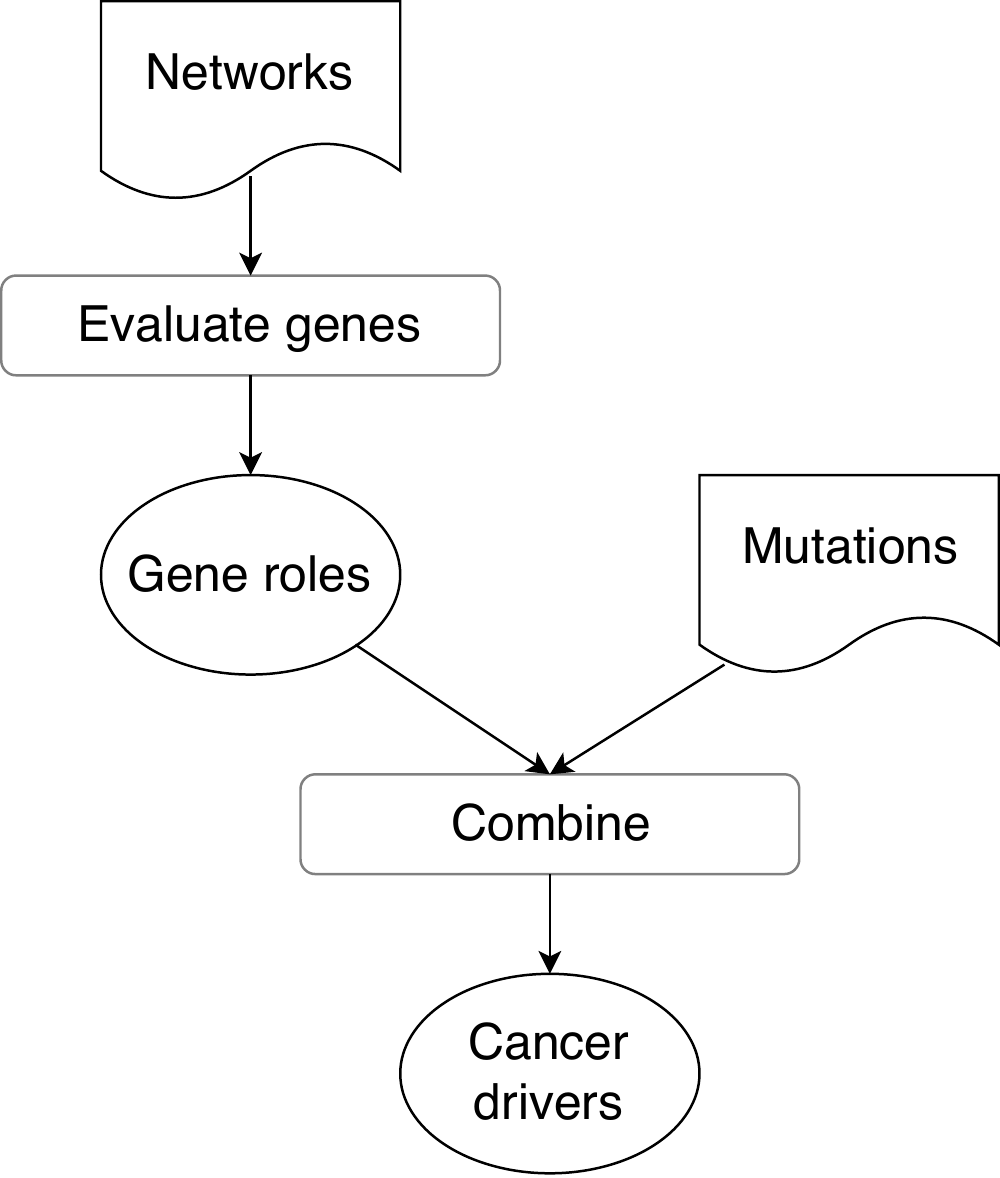}
\caption{Network-based methods. Network-based methods evaluate the role of genes in gene regulatory networks by using different techniques and combine with the mutations of genes to predict cancer drivers.}
\label{fig:UsingNetworks}
\end{figure}

Most current methods identify single cancer drivers at the population level. In general, they can be grouped in mutation-based methods and network-based methods. Mutation-based methods use the characteristics of mutations (e.g. the significance of mutations in genes, the functional impacts of mutations, the recurrence of mutations in genes, etc.) to identify cancer driver genes while network-based methods use gene networks to assess the role of genes then combine with the mutation information to predict cancer drivers. The general idea of the network-based methods is illustrated in Figure \ref{fig:UsingNetworks}.

\subsubsection{Mutation-based methods}

Mutation-based methods use the characteristics of mutations in genes to discover cancer driver genes. Based on the characteristics of mutations used in the methods, we divide them into four sub-groups, including using the significance of mutations in genes, using the functional impacts of mutations, using the recurrence of mutations, and others. Other methods combine the mutation information of genes with gene expression and tumour pathways to detect cancer drivers. The details of methods in the four sub-groups are discussed as below.

\textbf{A. Using the significance of mutations in genes}

MutSigCV \citep{RN233} is a method to discover cancer drivers by assessing the significance of mutations in genes. Cancer drivers predicted by MutSigCV are mutated more frequently than expected by chance based on inferred background mutation processes. However, MutSigCV has a limitation since although some genes have a high degree of mutations, these mutations are passenger mutations and do not contribute to the cancer development.

\textbf{B. Using the functional impacts of mutations}

OncodriveFM \citep{XRN245} uses the functional impact of genomic mutations to detect cancer drivers instead of evaluating the significance of mutations in genes like MutSigCV. OncodriveFM hypothesises that any bias of variations (i.e. mutations) in genes with a significantly functional impact may be an indicator for identifying candidate driver genes. The significant point of this method is that instead of assessing how many mutations a gene has, it evaluates how biased mutations with highly functional impacts are. Thus, it can detect driver genes having mutations with low recurrence but their mutations play a significant role in the cancer development.


Similar to OncodriveFM, OncodriveFML \citep{XRN344} also uses the functional impact of mutations to discover cancer drivers. However, while OncodriveFM only uses coding gene mutations, OncodriveFML is designed to analyse both coding and non-coding mutations. The OncodriveFML framework is then applied to 19 tumour datasets and uncovers well-known coding drivers like \textit{TP53, KEAP1, ARID2}, and \textit{RUNX1} with high functional impacts. It also identifies non-coding drivers such as \textit{MALAT1} and \textit{MIAT}. In particular, \textit{MALAT1} is a lncRNA which has been proved to be involved in lung adenocarcinomas and \textit{MIAT} is a non-protein-coding transcript related to myocardial infarction.

Another method assessing the functional impact of gene mutations to unravel cancer driver is DriverML \citep{RN232}. Different from OncodriveFM and OncodriveFML, DriverML assumes that the functional impact of mutations is affected by mutation types. Thus, it proposes a method to detect cancer drivers by scoring functional influences of alterations based on mutation types. The method uses various properties to weight the impact of mutation types and it obtains optimised weight parameters by using a supervised machine learning approach with pan-cancer training data.

\textbf{C. Using the recurrence of mutations in genes}

Instead of using the functional impact of mutations like OncodriveFM, OncodriveFML, and DriverML, other methods identify cancer drivers based on the location of mutations such as ActiveDriver \citep{RN234} and OncodriveCLUST \citep{XRN246}. ActiveDriver discovers cancer driver genes by detecting the enrichment of somatic mutations in post-translationally modified sites, including phosphorylation, acetylation, and ubiquitination sites. OncodriveCLUST is based on the fact that gain-of-function mutations usually cluster in particular protein sections and these mutations contribute to the development of cancer cells. Thus, it detects cancer genes with a large bias in clustering mutations. The method is applied to the database of Catalogue of Somatic Mutations in Cancer (COSMIC) \citep{XRN269} and then the result is validated with the Cancer Gene Census (CGC) \citep{XRN270}. As this method bases on the mutation clustering, it cannot identify cancer drivers whose mutations are distributed across the sequence. In addition, to have a good result, it requires a large number of observed mutations. Thus, this method should be used to complement results of other methods in detecting cancer drivers.


\textbf{D. Others: Combining with gene expression, pathways, etc.}

The platform IntOGen-mutations \citep{XRN234} is developed based on OncodriveFM and OncodriveCLUST to discover cancer drivers for various tumour types. This platform uses somatic mutations, gene expression, and tumour pathways as the input parameters. It takes the advantages of both methods using the functional impact of mutations and methods using the location of mutations by applying OncodriveFM to identify driver genes which are biased significantly toward mutations with high functional impacts and applying OncodriveCLUST to detect driver genes which have mutations highly concentrating in specific regions of proteins.


Also using mutational infomation in detecting cancer genes, PathScan \citep{XRN253} combines mutations with the information of genes in known pathways. PathScan tests the scenario in which pathway mutations contribute to the development of tumour. In addition, other methods combine mutations with existing knowledge of gene function or network structure, or finding mutually exclusive mutations, etc. For instance, Sakoparnig et al. \citep{XRN257} introduce a computational method to detect genomic alterations with low occurrence frequencies based on mutation timing.



Especially, some methods combine a wide range of data types in order to identify cancer drivers more effectively. For example, in \citep{XRN226}, the authors develop a computational framework which uses CNVs and gene expression as the inputs to uncover cancer drivers. The framework is named COpy Number and EXpression In Cancer (CONEXIC). It applies a score-guided search to detect combinations of modulators which reflect the expression of a gene module in a set of tumour samples. Then it identifies those having the highest score in amplified or deleted regions on chromosome. The authors hypothesise that in case the expression of gene \textit{A} and its copy number are related, the copy number variation likely results in changes in expression of gene \textit{A} and there is a high probability that \textit{A} is a driver candidate and it regulates other genes. The authors apply this framework to the dataset of melanoma and detect exactly its known cancer drivers.


ncDriver \citep{XRN238} identifies non-coding cancer drivers with a two-stage procedure. The first stage is mutational recurrence test which uses mutations (including indels and SNVs) and genomic elements as the inputs to detect elements with mutational recurrence. The second stage is to assess whether mutations of each element have a significant cancer-specific distribution and significant bias for highly conserved positions of each element, then it finds out if the conservation level of mutations is significantly large comparing to the overall conservation distribution. This procedure is applied to the pan-cancer whole-genome dataset to identify cancer drivers and significant non-coding drivers identified by the method are \textit{MIR142} lncRNA and \textit{XRNU5A-1} sncRNA.

\textbf{E. Analysis}

Although all the methods above base on mutation data to identify cancer drivers, each has a different approach. MutSigCV evaluates the significance of mutations in genes to detect cancer drivers. However, some genes are mutated significantly but most of their mutations are passenger mutations, which do not progress cancer. Thus, these genes are not cancer driver genes. To eliminate passenger mutations, ActiveDriver and OncodriveCLUST consider the location of mutations. Although these methods can reduce the false positives in predicting driver mutations, they may overlook cancer drivers with mutations distributing across the protein since they only evaluate mutations which are concentrated in particular protein sections. Instead of using the location of mutations, other methods use different strategies. For instance, OncodriveFM, OncodriveFML, and DriverML utilise the functional impact of genomic mutations to evaluate the importance of mutated genes to discover cancer drivers. Sakoparnig et al. \citep{XRN257} bases on the timing of mutations, PathScan combines with the pathway data, and CONEXIC combines with the gene expression data. There are also methods which use an integrated approach such as IntOGen-mutations, which considers both the functional impact of mutations and their clustering as well. Furthermore, since mutations in both coding regions and non-coding regions play a significant role in cancer development, cancer drivers can be coding or non-coding elements. Some methods like OncodriveFML and ncDriver are developed to detect non-coding cancer drivers.

As these methods evaluate different aspects of mutations to identify cancer drivers, they can detect several validated cancer drivers. The novel cancer drivers identified by these methods are potential and they can be used in wet-lab experiments to confirm their role in cancer progression. However, although these methods can be easily applied to different mutation datasets, mutation databases are incomplete and the applications of these methods are limited. 



\subsubsection{Network-based methods}

In general, network-based methods evaluate the role of genes in biological networks and then combine with the muttaion information of genes to predict cancer drivers. There are three methods in this group, including Vinayagam et al. \citep{XRN165}, CBNA \citep{RN228}, and DriverNet \citep{RN241}. The details of these methods are discussed as below.

\textbf{A. The details of methods}

Vinayagam et al. \citep{XRN165} applies controllability analysis on the directed network, i.e., the network with directed edges, of human protein-protein interaction (PPI). The input network includes nodes which are proteins and edges which are interactions between proteins. The controllability analysis categorises nodes into the three types which are "indispensable", "dispensable", or "neutral" based on their impact on minimum driver node set (MDS), i.e., the minimum node set driving the whole network. Indispensable nodes are nodes which make the number of MDS increased when the nodes are removed from the network, while dispensable nodes make the number of MDS decreased. The removal of neutral nodes from the network has no effect on the number of driver nodes. Then the study analyses the controllability of perturbated network to identify sensitive indispensable nodes, i.e., indispensable nodes in the original network but not in the perturbated network. These sensitive indispensable nodes are the candidate cancer drivers.

Also inspired by the network controllability, CBNA \citep{RN228} analyses the controllability of a gene regulatory network to discover cancer drivers. However, the network built by CBNA is a miRNA-TF-mRNA network which consists of microRNAs (miRNAs), Transcription Factors (TFs), and mRNAs. Since this network is constructed from the expression data of miRNAs/mRNAs of cancer patients and the existing gene interaction databases such as PPI \citep{RN235}, miRTarBase \citep{RN236}, and TransmiR \citep{RN240}, it is more reliable and specific to a cancer type. In addition, different from the method of Vinayagam et al. \citep{XRN165}, CBNA analyses the network controllability to indicate critical nodes of the network, i.e. nodes increase the number of the minimum node set controlling the whole network if they are removed from the network, then combining with the mutation data to identify cancer drivers. As CBNA uses the miRNA-TF-mRNA network, it can identify both coding and miRNA driver genes. Furthermore, it can also be used to discover drivers for a cancer type or cancer subtype.


Instead of evaluating the controllability of a subset of nodes of a gene network like Vinayagam et al. \citep{XRN165} and CBNA \citep{RN228}, DriverNet \citep{RN241} considers the influence of mutated genes on other genes in a network. DriverNet integrates different data types, including genome data (i.e. non-synonym SVNs, indels, and copy number variation), influence graph of biological pathway information, and gene expression. It constructs a bipartite graph of genes to detect the effect of mutated genes on genes which have an outlying expression. The putative drivers are mutated genes which impact on a high number of outlying-expression genes in several patients. The method is applied to four cancer datasets, including glioblastoma, breast, triple negative breast, and serous ovarian, and it reveals various candidate cancer drivers related to transcriptional networks.

\textbf{B. Analysis}

The three methods above use biological networks to predict single cancer drivers, other methods using networks to discover cancer driver modules or personalised cancer drivers are discussed in Section \ref{module} and \ref{personalised} respectively. In general, network-based methods evaluate the role of genes in the whole networks to predict cancer drivers. Various techniques are used to analyse the networks such as network controllability in Vinayagam et al. and CBNA or the influence of genes in DriverNet. These methods can elucidate molecular mechanisms in cancer development at the network level, but they need large datasets to produce reliable results. In addition, the networks used in some methods (i.e. Vinayagam et al. and DriverNet) are not specific to any cancer type, thus they may miss some important information which is specific to a cancer type. Another limitation of network-based methods like DriverNet is predicting genes which affect other genes' expression as cancer drivers, because some cancer drivers may not alter the expression of other genes or other genes accidentally change other genes' expression although they are not cancer drivers.




\subsection{Cancer driver module identification} \label{module}


Recently, several methods have been developed to discover cancer drivers in modules. Most of the methods identifying cancer driver modules use mutual exclusivity of mutations. Thus, we divide methods for identifying cancer driver modules into two sub-groups: using mutual exclusivity of mutations and others. Other methods use mutations, gene expression, gene network, etc. to detect cancer driver modules. The details of methods in the two sub-groups are discussed as below.

\textbf{A. Using mutual exclusivity of mutations}

CoMEt (the Combinations of Mutually Exclusive Alterations) \citep{XRN235} uses mutual exclusivity technique to detect cancer driver modules. Because different cancer patients have different combinations of genomic alterations which develop the disease, CoMEt detects combinations of alterations (i.e. modules of mutated genes) in the same pathway, which are mutual exclusive across samples. The method uses the exact statistical test to test mutual exclusivity and it does simultaneous analysis for mutually exclusive alterations specific to cancer subtypes. The advantage of this method is that it has a low computational complexity. Similarly, WeSME \citep{RN242} also assesses the mutual exclusivity of mutations of genes to detect cancer drivers. However, instead of evaluating genes in the same pathway, WeSME only considers gene pairs and the gene pairs whose mutations have a significantly mutual exclusivity are considered as modular candidate cancer drivers.

MEMo (Mutual Exclusivity Modules) \citep{XRN252} applies mutual exclusivity technique in biological networks to identify oncogenic network modules. According to \citep{XRN252}, although individual tumours of the same cancer type may have different genomic alterations, these alterations just happen in a restricted number of pathways. In addition, alterations in the same pathway are not likely to exist in the same patient. Based on these, MEMo does correlation analysis and applies statistical tests to detect network modules based on three criteria: (1) genes in a network module are altered across the sample; (2) member genes tend to join into the same biological process; (3) alterations in modules are mutually exclusive. The method is applied to the glioblastoma multiforme (GBM) dataset and detects successfully known network modules, i.e., groups of cancer drivers, in GBM.

\textbf{B. Others: Using mutations, gene expression, gene network, etc.}

iMCMC (an approach to identify Mutated Core Modules in Cancer) \citep{XRN228} is developed to uncover groups of genes driving cancer using the cancer genomic data from cancer patients. The method uses somatic mutation, CNV, and gene expression to build a gene network. Then, it identifies coherent subnetworks (modules) from the network through an optimisation model by selecting vertices and edges with high weights. Finally, the significance of subnetworks is assessed by performing a random test and the mutual exclusivity of subnetworks is tested by adopting Markov chain Monte Carlo permutation strategy. The method is applied to the GBM and the ovarian carcinoma (OV) datasets from TCGA. Many discovered core modules are related to known pathways and most of the identified genes are cancer driver genes which are already reported relating to cancer pathogenesis in other research.

NetBox \citep{XRN255} uses biological networks in studying drivers for GBM. It introduces a network-based method to detect oncogenic processes and cancer driver genes. The hypothesis of the approach is that biological networks include multiple functional modules, and tumours target specific functional modules. The method analyses sequence mutations, CNVs, an interaction network including both PPIs and signalling pathways to identify and assess network modules statistically.

Another method to identify cancer driver modules is TieDIE (Tied Diffusion Through Interacting Events) \citep{XRN254}. TieDIE applies network diffusion to discover the relationship of genomic events and changes in cancer subtypes. The approach collects a subnetwork of PPIs, interactions of genomic perturbations, predicted transcription factor-to-target connections, and transcriptomic states from literature. The method is applied to the breast adenocarcinoma (BRCA) dataset of TCGA and it detects signalling pathways and interlinking genes corresponding to cancer signalling.


The methods above identify coding cancer driver modules. However, because non-coding RNAs (e.g. miRNAs) can modulate tumorigenesis by promoting or suppressing specific genes and various cancer types have overlaps in oncogenic pathways, a group of miRNAs which drives or suppresses tumorigenesis in different tumour types may exist. Hamilton et al. \citep{XRN236} use the pan-cancer dataset of TCGA and the miRNA target data of Argonaute Crosslinking Immunoprecipitation (AGO-CLIP) \citep{XRN273, XRN271, XRN272} to detect pan-cancer miRNA drivers. The idea is that the set of cancer miRNA drivers will modulate tumorigenesis and share a central core seed motif. The result shows that an oncogenic miRNA superfamily, which includes \textit{miR-17, miR-18, miR-19, miR-93, miR-130, miR-210}, and \textit{miR-455}, coregulates tumour suppressors through a \textit{GUGC} core motif.

\textbf{C. Analysis}

As can be seen from the methods above, most of the methods use mutual exclusivity of mutations to identify cancer driver modules. With this technique, the mutation from only one member in an identified module is enough to trigger cancer progression \citep{RN242}. Thus, the identified drivers in a module may not work together to regulate their targets to drive cancer. However, as discussed above, genes should collaborate to increase their influence on target genes to progress cancer. Therefore, it is necessary to develop novel methods to discover cancer driver groups whose members work in concert to initialise and develop cancer.


\subsection{Personalised cancer driver identification} \label{personalised}

The methods discussed in Section \ref{single} and \ref{module} discover cancer drivers at the population level. Since different patients possess different genomes and their diseases might be driven by different driver genes, it is necessary to investigate cancer drivers which are specific to an individual patient (i.e. personalised cancer drivers). There are three methods in this group, including DawnRank \citep{XRN237}, SCS \citep{RN243}, and PNC \citep{RN244}. All of them base on gene regulatory networks to predict personalised cancer drivers. The details of these methods are discussed as below.

\textbf{A. The details of methods}

A representative of methods for identifying personalised cancer drivers is DawnRank \citep{XRN237}. In general, the idea of the method is that mutations in genes which have higher connectivity in an interaction network are more impactful. DawnRank uses the information of gene expression and gene network as the inputs. In particular, it is a ranking framework which applies PageRank \citep{XRN261, XRN260} to evaluate the impact of genes on the gene network. The impact is presented in terms of network connectivity and the number of downstream genes expressed differentially. The higher the rank of a gene is, the more downstream genes it has effects on in the gene network. Ranks of genes are then combined with somatic alteration data like copy number variations to detect driver alterations. Although DawnRank bases on the same gene regulatory network for all patients, it assesses the impact of genes in each patient using the patient's gene expression data to detect personalised cancer drivers. The algorithm has been applied to TCGA datasets and it shows an effectiveness in detecting cancer drivers.

To assess the impact of genes in each patient, DawnRank uses the gene expression data of each patient, but it bases on the same gene regulatory network of all patients. As a result, it may miss important information of gene regulation of each patient. Thus, to detect personalised cancer drivers, SCS \citep{RN243} builds a gene regulatory network for each patient from the patient's gene expression data and its neighbour's gene expression data (i.e. the corresponding normal sample's gene expression data). SCS detects cancer driver genes as the minimal set of mutated genes which impacts on the maximal differentially expressed genes. Like SCS, PNC \citep{RN244} also uses the gene expression data of a patient and its neighbour to construct personalised networks. Nevertheless, PNC only selects edges which are different between the tumour and normal state. It then converts the gene regulatory network to a bipartite graph in which, nodes on the top represent genes and nodes on the bottom represent edges. PNC predicts cancer driver genes as the minimum gene set on the top of the bipartite graph which covers all the edges on the bottom.

\textbf{B. Analysis}

Although these methods can discover personalised cancer drivers, they still have some limitations. DawnRank bases on the same gene network of all patients. It ignores the network information specific to an individual patient, leading to false positives in its results. On the other hand, SCS and PNC use the genetic data of each patient to construct personalised gene networks. However, they require the genetic data of a pair of samples (i.e. a tumour and its tumour neighbour), but identifying the neighbour of a tumour is challenging and it is not always existing. In addition, these methods only discover coding cancer drivers while non-coding genes (e.g. miRNAs) can also be cancer drivers as discussed above.


\section{Resources for cancer driver research} \label{section:resources}

\begin{table*}[!ht]
\processtable{Summary of resources for identifying cancer drivers and validating the predicted results \label{Tab:03}} {\begin{tabular}{llll}\toprule 
Category&Resource&Website & Description\\\midrule
\specialcell[t]{For identifying\\cancer drivers}
& TCGA \citep{XRN333}&\specialcell[t]{\href{https://www.cancer.gov/about-nci/organization/ccg/research/structural-genomics/tcga}{https://www.cancer.gov/about-nci/organization/\\ccg/research/structural-genomics/tcga}}& \specialcell[t]{Profiles human tumours to discover molecular aberrations in\\DNA, XRNA, protein, and epigenetic levels.}\\
& ICGC \citep{XRN276}&\specialcell[t]{\href{https://icgc.org/}{https://icgc.org/}} & \specialcell[t]{A data portal of cancer gemomics of 50 cancer types.}\\
& cBioPortal \citep{XRN278}&\specialcell[t]{\href{http://www.cbioportal.org/}{http://www.cbioportal.org/}} & \specialcell[t]{A web interface for accessing cancer genomic data and\\analysing the data.}\\
& Cancer3D \citep{XRN285}&\specialcell[t]{\href{http://cancer3d.org/search}{http://cancer3d.org/search}} & \specialcell[t]{Contains mutations of more than 14,700 proteins and they are\\mapped to over 24,300 proteins of the Protein Data Bank \citep{XRN286}.}\\
& CCLE \citep{XRN287}&\specialcell[t]{\href{https://portals.broadinstitute.org/ccle}{https://portals.broadinstitute.org/ccle}} & \specialcell[t]{Includes SNVs, CNAs, and gene expression.}\\
& COSMIC \citep{XRN279}&\specialcell[t]{\href{https://cancer.sanger.ac.uk/cosmic}{https://cancer.sanger.ac.uk/cosmic}} & \specialcell[t]{Contains cancer mutations, including manually curated expert\\data and data from sequencing projects.}\\
\specialcell[t]{For validating\\predicted results}
& CGC \citep{XRN270}&\specialcell[t]{\href{https://cancer.sanger.ac.uk/census}{https://cancer.sanger.ac.uk/census}} & \specialcell[t]{Provides a list of cancer genes, which has been well\\established for cancer development.}\\
& AGCOH \citep{XRN281}&\specialcell[t]{\href{http://atlasgeneticsoncology.org/}{http://atlasgeneticsoncology.org/}} & \specialcell[t]{Contains about 1,500 cancer genes merged from numerous\\collaborative projects.}\\
& NCG \cite{XRN282}&\specialcell[t]{\href{http://ncg.kcl.ac.uk/}{http://ncg.kcl.ac.uk/}} & \specialcell[t]{Comprises more than 500 known cancer genes and over 1,000\\candidate cancer genes.}\\
& DGIdb \citep{XRN284}&\specialcell[t]{\href{http://www.dgidb.org/}{http://www.dgidb.org/}} & \specialcell[t]{Includes cancer drivers and drug-gene interactions.}\\
& OncomiR  \citep{XRN295}&\specialcell[t]{\href{http://www.oncomir.org/}{http://www.oncomir.org/}} & \specialcell[t]{A web interface for investigating miRNA dysregulation in\\cancer.}\\
\botrule
\end{tabular}}{}
\end{table*}

There are two types of resources for developing computational methods for cancer driver discovery, including input data to a method and resource for validation. As input, it can be gene expression data, network data, mutation data, etc. For validation, it can be a database with ground truth or partial ground truth. The resources are summarised in Table~\ref{Tab:03}.

For input data, several databases have been developed from cancer sequencing projects and they provide rich data used in cancer driver identification methods. TCGA \citep{XRN333} is a significant project in this area. The TCGA project profiles and analyses human tumours to uncover molecular aberrations in DNA, XRNA, protein, and epigenetic levels \citep{XRN333}. TCGA data can be accessed through the Genomic Data Commons (GDC) data portal \citep{XRN275}. ICGC data portal is also a resource for cancer genomics data and it contains the data of genomic abnormalities of 50 cancer types \citep{XRN276}. Another data portal for cancer genomics is cBioPortal \citep{XRN278}, which provides a web interface for accessing cancer genomic datasets, as well as for analysing and visualising the data online.

There are also some other resources which can be used for cancer driver discovery such as the Cancer3D \citep{XRN285}, the Cancer Cell Line Encyclopedia (CCLE) \citep{XRN287}, and the COSMIC database \citep{XRN279}. Cancer3D is a database which focuses on the influence of mutations on the structure of proteins and it provides the information for users to analyse distribution patterns of mutations and their relationship with changes in drug activity \citep{XRN285}. It contains mutations of more than 14,700 proteins, which are mapped to over 24,300 proteins in the Protein Data Bank \citep{XRN286}. The CCLE includes SNVs, CNAs, and gene expression \citep{XRN287}. The COSMIC database is a large and comprehensive source for investigating the mutational impact in cancer. It contains records of cancer mutations including both manually curated expert data and data from sequencing projects like TCGA or ICGC \citep{XRN279, XRN280}. It has more than two million coding point mutations and over six million non-coding mutations \citep{XRN279}.

For validating identified cancer drivers, several databases can be used currently like CGC \citep{XRN270} in the COSMIC database. The CGC provides a gene list which has been well established for cancer progression. This list was collected through a census of genes which are mutated or implicated causally in cancer progression \citep{XRN270}. These genes are also called cancer genes. Beside CGC in COSMIC, there are several sources which can be used for validating cancer drivers. The Atlas of Genetics and Cytogenetics in Oncology and Haematology (AGCOH) is another source for this purpose \citep{XRN281}. It comprises around 1,500 cancer genes which are merged results from numerous collaborative projects \citep{XRN281}. The Network of Cancer Genes (NCG) is an online database of cancer genes with over 500 known cancer genes and more than 1,000 candidate cancer genes \cite{XRN282}. Known cancer genes are genes which have already been confirmed through experiments while candidate cancer genes are those using statistical methods. One more database about disease genes is the Drug-Gene Interaction database (DGIdb) \citep{XRN284}. It contains not only cancer drivers but also the information about drugs and drug-gene interactions \citep{XRN284}.

At the present, while coding drivers are well established in cancer research, non-coding drivers are not. In \citep{XRN295}, the authors have recently introduced OncomiR, which is a resource for investigating miRNA dysregulation in cancer through a web interface. It does statistical analyses based on RNA-seq, miRNA-seq, and clinical information from TCGA to discover miRNAs which are related to cancer progression. Although this database may not be used as a ground truth to validate miRNA cancer drivers, it can be used as a channel to explore miRNA dysregulation in detecting miRNA cancer drivers. To validate non-coding cancer drivers now, it is required to examine the literature manually \citep{XRN288, XRN294}. 


\section{Comparative study} \label{section:casestudy}

\subsection{Performance of methods in identifying cancer drivers} 

In this section, we present a comparative study to compare the performance of some methods above. As there is not a ground truth to compare the results of methods for discovering cancer driver modules, we only select five methods for identifying single cancer drivers and three methods for identifying personalised cancer drivers for the comparison, including ActiveDriver \citep{RN234}, DawnRank \citep{XRN237}, DriverML \citep{RN232}, DriverNet \citep{RN241}, MutSigCV \citep{RN233}, OncodriveFM \citep{XRN245}, PNC \citep{RN244}, and SCS \citep{RN243}. These methods represent for different approaches in detecting cancer driver genes. ActiveDriver, DriverML, MutSigCV, and OncodriveFM are mutation-based methods while DawnRank, DriverNet, PNC, and SCS are network-based methods. In addition, DawnRank, PNC, and SCS identify personalised cancer drivers while other five methods identify cancer drivers at the population level. Although DawnRank, PNC, and SCS detect cancer drivers for each patient, they all have a method to aggregate the results of individual patients to predict cancer drivers for the population. Thus, we can compare these three methods with the others. The comparison is performed based on the results of the eight methods in identifying drivers for breast invasive carcinoma (BRCA), lung adenocarcinoma (LUAD), lung squamous cell carcinoma (LUSC), kidney renal clear cell carcinoma (KIRC), head and neck squamous cell carcinoma (HNSC). We obtain the predicted cancer drivers of the eight methods for the selected five cancer types from \citep{RN244}.

The cancer drivers predicted by the methods are validated with the CGC from the COSMIC database as this database has catalogued the confirmed cancer drivers. The performance of a method is measured using $F_1 Score$ based on the number of discovered cancer drivers that are validated by the CGC. The $F_1 Score$ indicates the enrichment ability of discovered cancer drivers in the gold standard (i.e. the CGC) and it is computed based on Precision $P$ and Recall $R$ as shown in Eq. \ref{eq:0301}. The higher the $F_1 Score$ a method has, the better the method is.

\begin{equation}
F_1 Score=2*\frac{P*R}{ P+R}.\label{eq:0301}
\end{equation}

In Eq. \ref{eq:0301}, $P$ (Precision) shows the fraction of predicted driver genes in the CGC among the predicted driver genes and $R$ (Recall) indicates the fraction of predicted driver genes in the CGC among the driver genes in the CGC. As $F_1 Score$ is computed from Precision $P$ and Recall $R$, it will indicate both the ability to predict exactly cancer drivers and the ability to predict many confirmed cancer drivers of a method.

\begin{figure}[h!]
\centering
\includegraphics[width=0.5\textwidth]{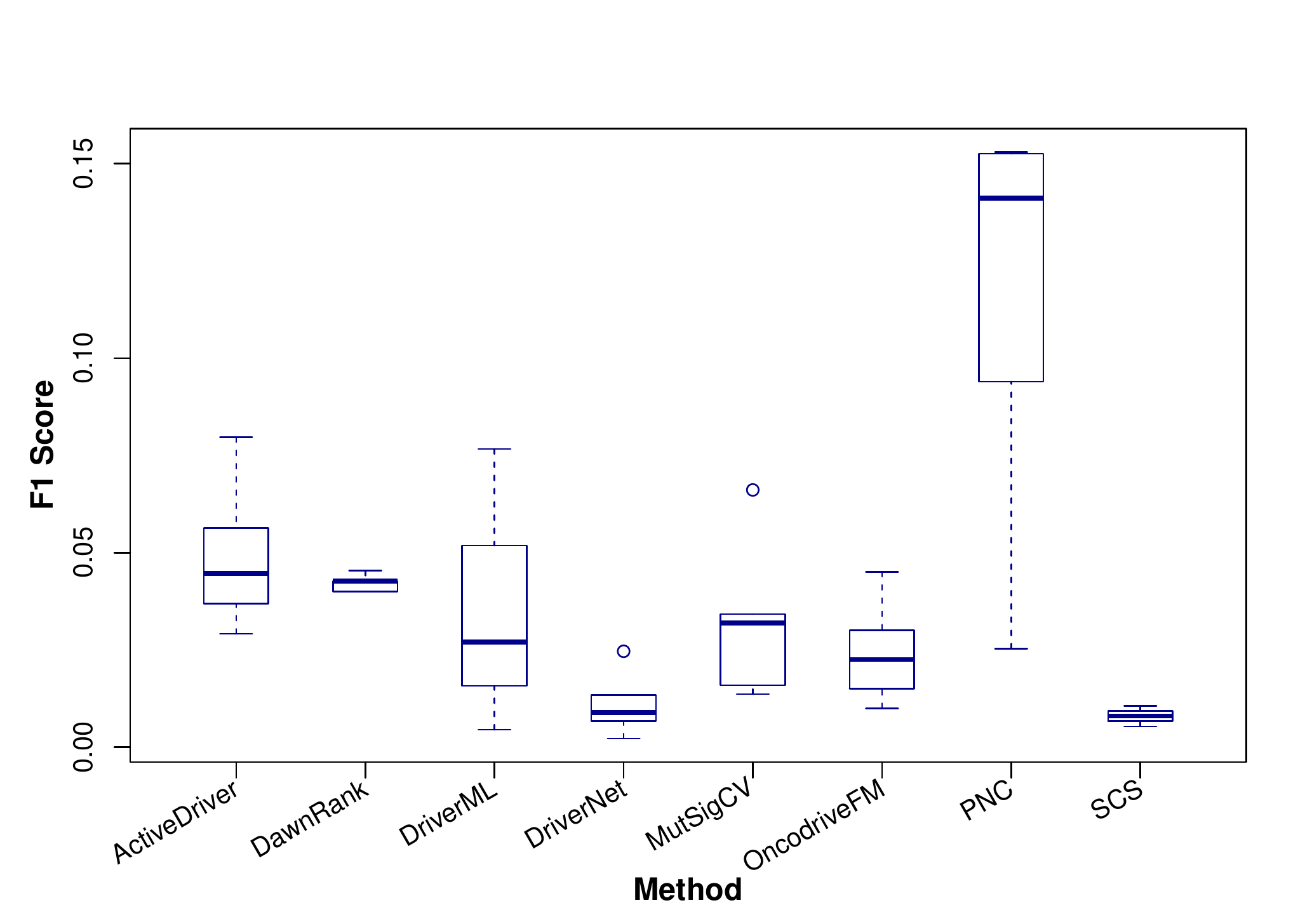}
\caption{Comparison of $F_1 Score$ of ActiveDriver, DawnRank, DriverML, DriverNet, MutSigCV, OncodriveFM, PNC, and SCS in identifying coding cancer drivers at the population level. The x-axis indicates the eight methods and the y-axis shows the $F_1 Score$. The results are based on the cancer driver prediction for the five cancer types, including BRCA, LUAD, LUSC, KIRC, and HNSC, of the eight methods.}
\label{fig:Comparison}
\end{figure}

The comparison result is shown in Figure~\ref{fig:Comparison} and the details are shown in Table~\ref{Tab:F1Socre}. It can be seen that with the four data sets of BRCA, LUAD, LUSC, and KIRC samples, PNC outperforms the other methods and with HNSC, ActiveDriver has the best performance.

\begin{table}[!ht]
\processtable{$F_1 Score$ of the eight methods in predicting drivers for the five cancer types \label{Tab:F1Socre}} {\begin{tabular}{clrrrrr}\toprule 
No. & Method & BRCA & LUAD & LUSC & KIRC & HNSC \\ \midrule
1 & ActiveDriver & 0.056 & 0.029 & 0.037 & 0.045 & 0.080 \\ 
  2 & DawnRank & 0.045 & 0.043 & 0.040 & 0.040 & 0.043 \\ 
  3 & DriverML & 0.077 & 0.027 & 0.016 & 0.052 & 0.005 \\ 
  4 & DriverNet & 0.007 & 0.009 & 0.013 & 0.025 & 0.002 \\ 
  5 & MutSigCV & 0.066 & 0.032 & 0.014 & 0.016 & 0.034 \\ 
  6 & OncodriveFM & 0.023 & 0.030 & 0.010 & 0.015 & 0.045 \\ 
  7 & PNC & 0.153 & 0.153 & 0.141 & 0.094 & 0.025 \\ 
  8 & SCS &NA  & 0.011 & 0.005 & 0.008 &NA  \\ 
\botrule
\end{tabular}}{}
\end{table}

\begin{figure*}[h!]
\centering
\includegraphics[width=1\textwidth]{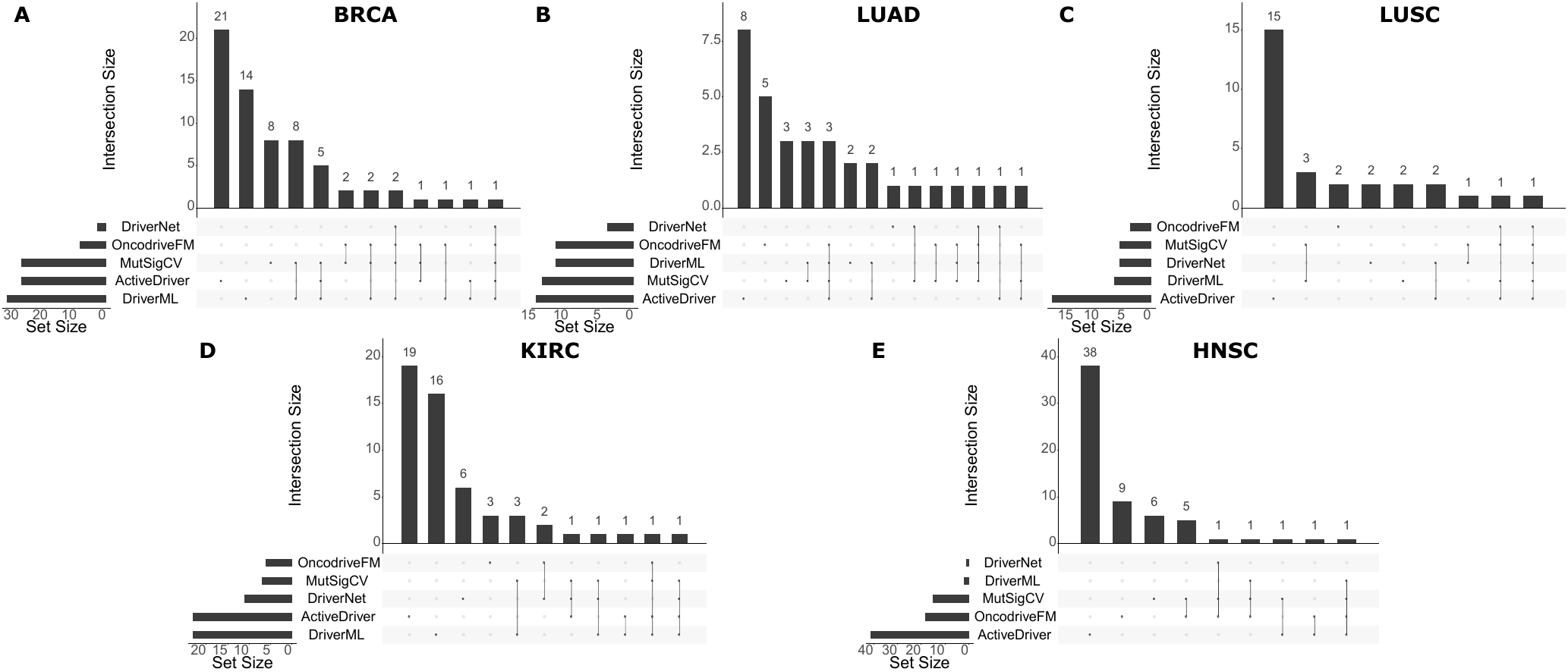}
\caption{Overlap among the cancer drivers predicted by different methods. The charts illustrate the overlap among the cancer drivers at the population level predicted by the five methods (DriverML, ActiveDriver, DriverNet, MutSigCV, and OncodriveFM) w.r.t the five cancer types, including BRCA, LUAD, LUSC, KIRC, and HNSC. In each chart, the horizontal bars at the bottom left show the number of detected cancer drivers validated by the CGC, the vertical bars and the dotted lines show the overlap of the validated cancer drivers of the methods. If there is not an overlap, it will be a black dot.}
\label{fig:Overlap}
\end{figure*}

Moreover, to see if the methods detect similar cancer drivers, we compare the results of the five methods used for identifying cancer drivers at the population level (i.e. DriverML, ActiveDriver, DriverNet, MutSigCV, and OncodriveFM).  Figure~\ref{fig:Overlap} shows the overlap between the validated cancer drivers  discovered by each pair of the methods, for each of the five cancer types. It can be seen that there is little overlap among the results of the methods. For example, in breast cancer, only one cancer driver (\textit{TP53}) is identified by all the five methods, two cancer drivers (\textit{CDH1} and \textit{PIK3CA}) are detected by four methods (DriverML, DriverNet, MutSigCV, and OncodriveFM), and eight cancer drivers (\textit{GATA3, NCOR1, PTEN, ARID1A, FOXA1, PIK3R1, CTCF}, and \textit{ERBB2}) are predicted by three methods. As the results of these methods are complementary, they should be used together to maximize the overall performance of the cancer driver prediction. In addition, it should be pointed out that although the CGC is popular in validating cancer drivers in cancer research, it is incomplete in the sense that the database is constantly being updated when new cancer drivers come to light. Therefore, although some of the predicted cancer drivers cannot be validated with existing knowledge, they can be novel cancer drivers which is worth wet-lab experiments to confirm their roles in progressing cancer.

\subsection{Identified cancer drivers enriched significantly in GO biological processes and KEGG pathways} 

Taking breast cancer as an example, we combine all the breast cancer drivers predicted by the five methods at the population level (i.e. DriverML, ActiveDriver, DriverNet, MutSigCV, and OncodriveFM), which results in altogether 509 cancer drivers. Among them, 63 drivers are predicted by at least two of the five methods. We use Enrichr \citep{RN245} to do enrichment analysis of these 63 drivers. Table~\ref{Tab:GO} and Table~\ref{Tab:KEGG} show the GO biological processes and KEGG pathways in which these cancer drivers are significantly enriched (adjusted p-value < 0.05). Among the 63 driver genes, 16 genes (25.4\%) are enriched in 7 GO biological processes and 15 genes (23.8\%) are enriched in 26 KEGG pathways related to breast cancer. It indicates that the predicted cancer drivers are closely associated with the biological condition of breast cancer and biologically meaningful.


\begin{table}[!ht]
\processtable{GO biological processes involved in breast cancer in which the predicted cancer drivers are enriched \label{Tab:GO}} {\begin{tabular}{lrr}\toprule 
Term & \#Genes & p-value \\  \midrule
GO:0045598 regulation of fat cell differentiation &  5 & 2.0e-03 \\ 
GO:0045596 negative regulation of cell differentiation &  6 & 3.6e-03 \\ 
GO:0045604 regulation of epidermal cell differentiation&  3 & 1.2e-02 \\ 
GO:0042127 regulation of cell proliferation & 10 & 2.5e-02 \\ 
GO:0045599 negative regulation of fat cell differentiation &  3 & 2.8e-02 \\ 
GO:0045580 regulation of T cell differentiation &  3 & 2.9e-02 \\ 
GO:2000736 regulation of stem cell differentiation&  4 & 3.1e-02 \\ 
\botrule
\end{tabular}}{}
\end{table}

\begin{table}[!ht]
\processtable{KEGG pathways involved in breast cancer in which the predicted cancer drivers are enriched \label{Tab:KEGG}} {\begin{tabular}{lrr}\toprule 
Term & \#Genes & p-value \\  \midrule
ErbB signaling pathway &  6 & 5.3e-06 \\ 
Thyroid hormone signaling pathway &  6 & 2.8e-05 \\ 
Sphingolipid signaling pathway &  6 & 3.1e-05 \\ 
Neurotrophin signaling pathway &  6 & 3.0e-05 \\ 
PI3K-Akt signaling pathway &  8 & 1.7e-04 \\ 
AGE-RAGE signaling pathway in diabetic complications &  5 & 1.7e-04 \\ 
HIF-1 signaling pathway &  5 & 1.7e-04 \\ 
FoxO signaling pathway &  5 & 5.1e-04 \\ 
Fc epsilon RI signaling pathway &  4 & 5.2e-04 \\ 
Toll-like receptor signaling pathway &  4 & 2.2e-03 \\ 
TNF signaling pathway &  4 & 2.7e-03 \\ 
Relaxin signaling pathway &  4 & 4.6e-03 \\ 
VEGF signaling pathway &  3 & 5.1e-03 \\ 
Estrogen signaling pathway &  4 & 5.3e-03 \\ 
mTOR signaling pathway &  4 & 7.3e-03 \\ 
Prolactin signaling pathway &  3 & 7.4e-03 \\ 
B cell receptor signaling pathway &  3 & 7.6e-03 \\ 
p53 signaling pathway &  3 & 7.8e-03 \\ 
MAPK signaling pathway &  5 & 1.2e-02 \\ 
T cell receptor signaling pathway &  3 & 1.8e-02 \\ 
Rap1 signaling pathway &  4 & 1.8e-02 \\ 
C-type lectin receptor signaling pathway &  3 & 1.9e-02 \\ 
AMPK signaling pathway &  3 & 2.6e-02 \\ 
Apelin signaling pathway &  3 & 3.5e-02 \\ 
Insulin signaling pathway &  3 & 3.4e-02 \\ 
Phospholipase D signaling pathway &  3 & 4.1e-02 \\ 
\botrule
\end{tabular}}{}
\end{table}

\subsection{Identified cancer drivers are useful in predicting survival} 

\begin{figure}[h!]
\centering
\includegraphics[width=0.5\textwidth]{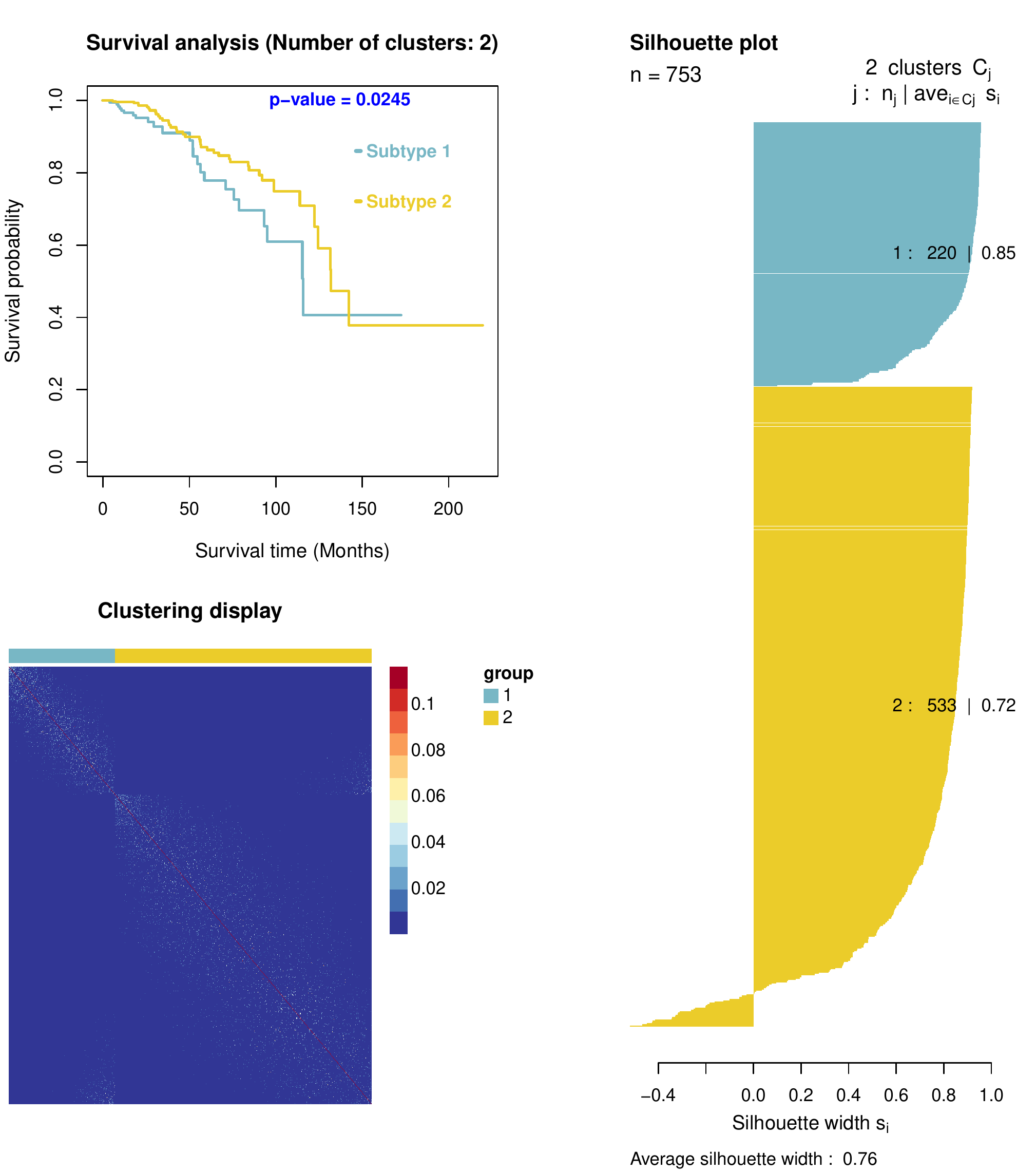}
\caption{Survival curves, clustering display, and silhouette plot. Survival curves are for cancer subtypes identified by using the four predicted cancer drivers, including \textit{AKT1, PTEN, CDKN1B}, and \textit{TP53}. The survival curves show the significant difference in the survivals of patients of the two subtypes (p-value = 0.0245). The clustering display indicates a highly qualified clustering with the similarity of samples in each subtype (i.e. Light dots show the similarity of samples). The silhouette plot has a large average silhouette width (0.76/1), indicating the clustering validity when using these four genes.}
\label{fig:Survival}
\end{figure}

Since the predicted cancer driver genes likely cause carcinogenesis, they could be used as biomarkers to classify tumours. To explore this concept, we use the predicted drivers to stratify breast cancer patients. Among the 63 predicted cancer drivers above, there are four significant genes, \textit{AKT1, PTEN, CDKN1B}, and \textit{TP53}, which are enriched in both GO biological processes and KEGG pathways. For instance, \textit{AKT1} are enriched in two GO biological processes and 25 KEGG pathways, \textit{PTEN} are enriched in two GO biological processes and five KEGG pathways. Thus, we use these four genes for this analysis. In addition, we obtain the BRCA gene expression data and clinical data from \citep{RN230}, and use the Similarity Network Fusion (SNF) method \citep{RN251, RN250}, a popular method for discovering the similarities among patients, to cluster cancer patients. The SNF takes expression of these four genes as input and outputs subtypes of cancer patients. We then analyse the survival outcomes of patients in the classified subtypes. The results indicate that the survival level of patients in different classified subtypes are significantly different (p-value = 0.0245) as shown in Figure~\ref{fig:Survival}. Furthermore, the clustering display shows the similarity of samples in each identified subtype and the silhouette plot indicates a good clustering with a large average silhouette width (0.76).

\section{Gaps and future directions} \label{section:gap}

From the discussion above, we see that there are a wide range of computational methods for identifying cancer drivers from genomic data. In this paper, we categorise the methods into three groups: methods for identifying single cancer drivers (including mutation-based methods and network-based methods), methods for identifying cancer driver modules, and methods for identifying personalised cancer drivers. Although these methods have detected successfully various cancer drivers, there are still several gaps in the research of the field.

Firstly, most of the current methods focus on coding mutations to identify coding cancer drivers while non-coding cancer drivers are not fully examined and the number of methods for identifying non-coding drivers is limited. However, non-coding cancer drivers are important because protein-coding regions account for only around two percent of the human genome. The large part of mutations exist in non-coding regions and these mutations can regulate the expression of genes and drive cancer \citep{XRN297, XRN298}. In addition to the limited number of non-coding cancer driver identification methods, the current methods focus much on non-coding mutations, i.e., correlations of mutations in non-coding elements with other factors like survival \citep{XRN238}. Nevertheless, cancer drivers can be non-coding RNAs without mutations but they can regulate other genes to progress cancer, thus it is required to investigate non-coding RNAs with and without mutations to detect non-coding cancer drivers.

Secondly, some methods have been developed to identify groups of cancer drivers \citep{XRN252, XRN228}, but they are mostly based on mutations to detect mutated modules, called cancer driver modules. Since in a module, the mutation of a member is sufficient to develop cancer, the identified drivers in a module may not in fact work together to regulate their targets to drive cancer. However, there is evidence that some genes work in concert to regulate other genes' expression and influence different biological processes, such as the cooperation of miRNAs in EMT, the transformation of epithelial cells into mesenchymal cells \citep{XRN326, XRN327}. In addition, in some biological processes, the regulation of single genes might not have significant impacts and research has emerged to use wet-lab experiments to investigate the regulatory of group-based regulators in biological processes. All of these highlight the importance of studying biological factors in groups, and computational methods which utilise a variety of data and techniques are in demand for investigating groups of cancer drivers.

Finally, although there have been methods for detecting personalised cancer drivers \citep{RN243, RN244, XRN237}, they still have some limitations. Some methods, such as DawnRank, use the gene network of the population to predict personlaised cancer drivers. This leads to that they may ignore the information of the gene network specific to an individual patient and they may discover many false positives in their results. Other methods, such as SCS and PNC, use the personal genetic data to build personalised gene networks but they need the genetic data of a sample pair (i.e. a cancer patient and its neighbour in the nomal state). The neighbour of a cancer patient is not always existing. Thus, the application of these methods is limited. Furthermore, these methods only detect coding cancer drivers while it is also necessary to identify non-coding cancer drivers as the discussion above. All of these indicate that there is a strong need to develop novel computational methods for detecting personalised coding/non-coding cancer drivers.


\section{Recommendation and conclusion} \label{section:conclusion}

We have investigated a wide range of computational methods for identifying cancer drivers from genomic data. In addition, the advantages and limitations of the surveyed methods are analysed, based on which we identify various opportunities for the development of the research in the field. It is clear that the research in computational approach to cancer driver identification is still in its growth phase. Much more work needs to be done and many opportunities exist in this area. Nevertheless, there are also different challenges in advancing the research in cancer driver identification. Identifying exactly biological factors which drive cancer is quite complicated. Future research needs to focus on both coding and non-coding datasets to identify candidate cancer drivers. To improve the accuracy of the novel computational methods, we should combine different types of data such as gene expression, mutations, and clinical information, etc. to detect cancer drivers.

We have also surveyed available resources which can be used in the research of discovering cancer drivers. The existing resources are plentiful but they are fragmented. Thus, to utilise cancer data more effectively for the research, it requires to have policies to achieve better data sharing. In addition, another difficulty when developing computational methods for uncovering non-coding cancer drivers is the validation. The reason is that most of the current databases are for coding cancer drivers and there is no one for non-coding cancer drivers. Therefore, we make an urgent call for the building of databases for non-coding drivers given their crucial role in the success of the research in the field.

To evaluate the performance of some current methods in detecting cancer drivers as well as provide an example of the evaluation of cancer driver discovery methods for the researchers who would like to penetrate the field, a comparative study has been conducted. From the results of the experiment in the comparative study, it can be seen that each method can uncover different cancer drivers and the overlaps between the results of the methods are small. Therefore, the methods are complementary, and we should use them together to maximize the effectiveness of cancer driver prediction of the methods. This is also an indicator for the different approaches of the methods and to achieve a significant result, novel methods should combine various resources and techniques in detecting cancer drivers.

In conclusion, although there are numerous computational methods for discovering cancer drivers now, there exist various gaps and opportunities for advancing the research of the field. However, due to the complexity of cancer initialisation and development, identifying cancer drivers faces many challenges. Through this paper, we hope that we can help researchers who are interested in the filed to establish a solid background and motivate them to tackle the current challenges.

\section*{Acknowledgements}

This research is supported by the Australian Government Research Training Program (RTP) Scholarship and the Vice Chancellor \& President's Scholarship offered by the University of South Australia.\vspace*{-12pt}

\section*{Funding}

The ARC DECRA (No: 200100200) and the Australian Research Council Discovery Grant (No: DP170101306).\vspace*{-12pt}

\section*{Biographical note}

\textbf{Vu Viet Hoang Pham} is a PhD student at UniSA STEM. He received his Master of Information Technology in 2017 at Deakin University. His research interests are causal inference and its applications in Bioinformatics.

\textbf{Lin Liu} is an associate professor at UniSA STEM. She received her bachelor and master degrees in Electronic Engineering from Xidian University, China in 1991 and 1994 respectively, and her PhD degree in computer systems engineering from UniSA in 2006. Her research interests include data mining, causal discovery and their applications in bioinformatics.

\textbf{Cameron Bracken} is a lab head at the Centre for Cancer Biology, an alliance between SA Pathology and University of South Australia. His research interests are the mechanisms that non-coding RNAs regulate EMT.

\textbf{Gregory Goodall} is a professor at Centre for Cancer Biology, an alliance of SA Pathology and University of South Australia. He is a world leader in the biology of RNA and cancer progression. He has combined innovation with thoroughness to make discoveries that open new areas in RNA biology for development and exploitation. He has made seminal contributions to the understanding of mechanisms governing gene activity in cancer, through control of mRNA activity, regulation of gene expression by microRNAs, and most recently his discovery of the regulation of circular RNAs. These breakthroughs have widespread implications for understanding gene regulation in biology, particularly in immunity and cancer. 

\textbf{Jiuyong Li} is a professor at UniSA STEM. He received his PhD degree in computer science from the Griffith University, Australia (2002). His research interests are in the fields of data mining, privacy preserving and bioinformatics. His research has been supported by six prestigious Australian Research Council Discovery grants since 2005 and he has published more than 100 research papers.

\textbf{Thuc Duy Le} is a senior lecturer at UniSA STEM. He is also an ARC DECRA fellow in Bioinformatics. He received his PhD degree in Computer Science (Bioinformatics) in 2014 at UniSA. His research interests are causal inference and its applications in bioinformatics.

\section*{Key points}

\begin{itemize}
  \item Providing a comprehensive survey of cancer driver discovery methods
  \item Categorising methods for identifying cancer drivers into three groups: methods for identifying single cancer drivers, methods for identifying cancer driver modules, and methods for identifying personalised cancer drivers
  \item Introducing several resources for cancer driver identification research
  \item Performing a case study to compare the performance of the current methods for identifying cancer drivers and analysing their results
  \item Analysing the advatages and limitations of the current methods as well as identifying the opportunities and challenges in developing reliable cancer driver discovery methods
\end{itemize}

 \bibliographystyle{natbib}
\bibliography{document}

\begin{thebibliography}{}

\bibitem[Akavia {\em et~al.}(2010)Akavia, Litvin, Kim, {\em et~al.}]{XRN226}
Akavia, U.~D., Litvin, O., Kim, J., {\em et~al.} (2010).
\newblock An integrated approach to uncover drivers of cancer.
\newblock {\em Cell\/}, {\bf 143}(6), 1005--1017.

\bibitem[An {\em et~al.}(2016)An, Dall'Olio, Mourikis, {\em et~al.}]{XRN282}
An, O., Dall'Olio, G.~M., Mourikis, T.~P., {\em et~al.} (2016).
\newblock Ncg 5.0: updates of a manually curated repository of cancer genes and
  associated properties from cancer mutational screenings.
\newblock {\em Nucleic Acids Research\/}, {\bf 44}(Database issue), D992--D999.

\bibitem[Barretina {\em et~al.}(2012)Barretina, Caponigro, Stransky, {\em
  et~al.}]{XRN287}
Barretina, J., Caponigro, G., Stransky, N., {\em et~al.} (2012).
\newblock The cancer cell line encyclopedia enables predictive modeling of
  anticancer drug sensitivity.
\newblock {\em Nature\/}, {\bf 483}(7391), 603--607.

\bibitem[Bashashati {\em et~al.}(2012)Bashashati, Haffari, Ding, {\em
  et~al.}]{RN241}
Bashashati, A., Haffari, G., Ding, J., {\em et~al.} (2012).
\newblock Drivernet: uncovering the impact of somatic driver mutations on
  transcriptional networks in cancer.
\newblock {\em Genome biology\/}, {\bf 13}(12), R124--R124.

\bibitem[Brin and Page(1998)Brin and Page]{XRN261}
Brin, S. and Page, L. (1998).
\newblock The anatomy of a large-scale hypertextual web search engine.
\newblock {\em Comput. Netw. ISDN Syst.}, {\bf 30}(1-7), 107--117.

\bibitem[Cerami {\em et~al.}(2010)Cerami, Demir, Schultz, {\em et~al.}]{XRN255}
Cerami, E., Demir, E., Schultz, N., {\em et~al.} (2010).
\newblock Automated network analysis identifies core pathways in glioblastoma.
\newblock {\em PLOS ONE\/}, {\bf 5}(2), e8918.

\bibitem[Chi {\em et~al.}(2009)Chi, Zang, Mele, {\em et~al.}]{XRN273}
Chi, S.~W., Zang, J.~B., Mele, A., {\em et~al.} (2009).
\newblock Ago hits-clip decodes mirna-mrna interaction maps.
\newblock {\em Nature\/}, {\bf 460}(7254), 479--486.

\bibitem[Chou {\em et~al.}(2016)Chou, Chang, Shrestha, {\em et~al.}]{RN236}
Chou, C.-H., Chang, N.-W., Shrestha, S., {\em et~al.} (2016).
\newblock mirtarbase 2016: updates to the experimentally validated mirna-target
  interactions database.
\newblock {\em Nucleic acids research\/}, {\bf 44}(D1), D239--D247.

\bibitem[Ciriello {\em et~al.}(2012)Ciriello, Cerami, Sander, {\em
  et~al.}]{XRN252}
Ciriello, G., Cerami, E., Sander, C., {\em et~al.} (2012).
\newblock Mutual exclusivity analysis identifies oncogenic network modules.
\newblock {\em Genome Research\/}, {\bf 22}(2), 398--406.

\bibitem[Cursons {\em et~al.}(2017)Cursons, Pillman, Scheer, {\em
  et~al.}]{XRN326}
Cursons, J., Pillman, K.~A., Scheer, K., {\em et~al.} (2017).
\newblock Post-transcriptional control of emt is coordinated through
  combinatorial targeting by multiple micrornas.
\newblock {\em bioRxiv\/}.

\bibitem[Cuykendall {\em et~al.}(2017)Cuykendall, Rubin, and Khurana]{XRN288}
Cuykendall, T.~N., Rubin, M.~A., and Khurana, E. (2017).
\newblock Non-coding genetic variation in cancer.
\newblock {\em Current Opinion in Systems Biology\/}, {\bf 1}, 9--15.

\bibitem[Dimitrakopoulos and Beerenwinkel(2017)Dimitrakopoulos and
  Beerenwinkel]{XRN239}
Dimitrakopoulos, C.~M. and Beerenwinkel, N. (2017).
\newblock Computational approaches for the identification of cancer genes and
  pathways.
\newblock {\em Wiley Interdisciplinary Reviews. Systems Biology and
  Medicine\/}, {\bf 9}(1), e1364.

\bibitem[Fattore {\em et~al.}(2016)Fattore, Mancini, Acunzo, {\em
  et~al.}]{XRN192}
Fattore, L., Mancini, R., Acunzo, M., {\em et~al.} (2016).
\newblock mir-579-3p controls melanoma progression and resistance to target
  therapy.
\newblock {\em Proceedings of the National Academy of Sciences\/}, {\bf
  113}(34), E5005.

\bibitem[Forbes {\em et~al.}(2010)Forbes, Tang, Bindal, {\em et~al.}]{XRN269}
Forbes, S.~A., Tang, G., Bindal, N., {\em et~al.} (2010).
\newblock Cosmic (the catalogue of somatic mutations in cancer): a resource to
  investigate acquired mutations in human cancer.
\newblock {\em Nucleic Acids Research\/}, {\bf 38}(Database issue), D652--D657.

\bibitem[Forbes {\em et~al.}(2011)Forbes, Bindal, Bamford, {\em
  et~al.}]{XRN280}
Forbes, S.~A., Bindal, N., Bamford, S., {\em et~al.} (2011).
\newblock Cosmic: mining complete cancer genomes in the catalogue of somatic
  mutations in cancer.
\newblock {\em Nucleic Acids Research\/}, {\bf 39}(Database issue), D945--D950.

\bibitem[Forbes {\em et~al.}(2015)Forbes, Beare, Gunasekaran, {\em
  et~al.}]{XRN279}
Forbes, S.~A., Beare, D., Gunasekaran, P., {\em et~al.} (2015).
\newblock Cosmic: exploring the world's knowledge of somatic mutations in human
  cancer.
\newblock {\em Nucleic Acids Research\/}, {\bf 43}(Database issue), D805--D811.

\bibitem[Futreal {\em et~al.}(2004)Futreal, Coin, Marshall, {\em
  et~al.}]{XRN270}
Futreal, P.~A., Coin, L., Marshall, M., {\em et~al.} (2004).
\newblock A census of human cancer genes.
\newblock {\em Nature reviews. Cancer\/}, {\bf 4}(3), 177--183.

\bibitem[Gao {\em et~al.}(2013)Gao, Aksoy, Dogrusoz, {\em et~al.}]{XRN278}
Gao, J., Aksoy, B.~A., Dogrusoz, U., {\em et~al.} (2013).
\newblock Integrative analysis of complex cancer genomics and clinical profiles
  using the cbioportal.
\newblock {\em Science Signaling\/}, {\bf 6}(269), pl1.

\bibitem[Gasparini {\em et~al.}(2015)Gasparini, Cascione, Landi, {\em
  et~al.}]{XRN177}
Gasparini, P., Cascione, L., Landi, L., {\em et~al.} (2015).
\newblock microrna classifiers are powerful diagnostic/prognostic tools in
  alk-, egfr-, and kras-driven lung cancers.
\newblock {\em Proceedings of the National Academy of Sciences\/}, {\bf
  112}(48), 14924.

\bibitem[Gonzalez-Perez and Lopez-Bigas(2012)Gonzalez-Perez and
  Lopez-Bigas]{XRN245}
Gonzalez-Perez, A. and Lopez-Bigas, N. (2012).
\newblock Functional impact bias reveals cancer drivers.
\newblock {\em Nucleic Acids Research\/}, {\bf 40}(21), e169--e169.

\bibitem[Gonzalez-Perez {\em et~al.}(2013)Gonzalez-Perez, Perez-Llamas,
  Deu-Pons, {\em et~al.}]{XRN234}
Gonzalez-Perez, A., Perez-Llamas, C., Deu-Pons, J., {\em et~al.} (2013).
\newblock Intogen-mutations identifies cancer drivers across tumor types.
\newblock {\em Nature Methods\/}, {\bf 10}, 1081.

\bibitem[Griffith {\em et~al.}(2013)Griffith, Griffith, Coffman, {\em
  et~al.}]{XRN284}
Griffith, M., Griffith, O.~L., Coffman, A.~C., {\em et~al.} (2013).
\newblock Dgidb: mining the druggable genome.
\newblock {\em Nature Methods\/}, {\bf 10}, 1209.

\bibitem[Grossman {\em et~al.}(2016)Grossman, Heath, Ferretti, {\em
  et~al.}]{XRN275}
Grossman, R.~L., Heath, A.~P., Ferretti, V., {\em et~al.} (2016).
\newblock Toward a shared vision for cancer genomic data.
\newblock {\em New England Journal of Medicine\/}, {\bf 375}(12), 1109--1112.

\bibitem[Guo {\em et~al.}(2018)Guo, Zhang, Liu, {\em et~al.}]{RN243}
Guo, W.-F., Zhang, S.-W., Liu, L.-L., {\em et~al.} (2018).
\newblock Discovering personalized driver mutation profiles of single samples
  in cancer by network control strategy.
\newblock {\em Bioinformatics\/}, {\bf 34}(11), 1893--1903.

\bibitem[Guo {\em et~al.}(2019)Guo, Zhang, Zeng, {\em et~al.}]{RN244}
Guo, W.~F., Zhang, S.~W., Zeng, T., {\em et~al.} (2019).
\newblock A novel network control model for identifying personalized driver
  genes in cancer.
\newblock {\em PLoS Comput Biol\/}, {\bf 15}(11), e1007520.

\bibitem[Hafner {\em et~al.}(2010)Hafner, Landthaler, Burger, {\em
  et~al.}]{XRN271}
Hafner, M., Landthaler, M., Burger, L., {\em et~al.} (2010).
\newblock Transcriptome-wide identification of rna-binding protein and microrna
  target sites by par-clip.
\newblock {\em Cell\/}, {\bf 141}(1), 129--141.

\bibitem[Hafner {\em et~al.}(2012)Hafner, Lianoglou, Tuschl, {\em
  et~al.}]{XRN272}
Hafner, M., Lianoglou, S., Tuschl, T., {\em et~al.} (2012).
\newblock Genome-wide identification of mirna targets by par-clip.
\newblock {\em Methods (San Diego, Calif.)\/}, {\bf 58}(2), 94--105.

\bibitem[Hamilton {\em et~al.}(2013)Hamilton, Rajapakshe, Hartig, {\em
  et~al.}]{XRN236}
Hamilton, M.~P., Rajapakshe, K., Hartig, S.~M., {\em et~al.} (2013).
\newblock Identification of a pan-cancer oncogenic microrna superfamily
  anchored by a central core seed motif.
\newblock {\em Nature Communications\/}, {\bf 4}, 2730.

\bibitem[Han {\em et~al.}(2019)Han, Yang, Qian, {\em et~al.}]{RN232}
Han, Y., Yang, J., Qian, X., {\em et~al.} (2019).
\newblock Driverml: a machine learning algorithm for identifying driver genes
  in cancer sequencing studies.
\newblock {\em Nucleic Acids Research\/}, {\bf 47}(8), e45--e45.

\bibitem[Hornshoj {\em et~al.}(2018)Hornshoj, Nielsen, Sinnott-Armstrong, {\em
  et~al.}]{XRN238}
Hornshoj, H., Nielsen, M.~M., Sinnott-Armstrong, N.~A., {\em et~al.} (2018).
\newblock Pan-cancer screen for mutations in non-coding elements with
  conservation and cancer specificity reveals correlations with expression and
  survival.
\newblock {\em npj Genomic Medicine\/}, {\bf 3}(1), 1.

\bibitem[Hou and Ma(2014)Hou and Ma]{XRN237}
Hou, J.~P. and Ma, J. (2014).
\newblock Dawnrank: discovering personalized driver genes in cancer.
\newblock {\em Genome Medicine\/}, {\bf 6}(7), 56.

\bibitem[Huret {\em et~al.}(2000)Huret, Minor, Dorkeld, {\em et~al.}]{XRN281}
Huret, J.-L., Minor, S.~L., Dorkeld, F., {\em et~al.} (2000).
\newblock Atlas of genetics and cytogenetics in oncology and haematology, an
  interactive database.
\newblock {\em Nucleic Acids Research\/}, {\bf 28}(1), 349--351.

\bibitem[Institute(2018)Institute]{XRN333}
Institute, N. H. G.~R. (2018).
\newblock The cancer genome atlas.

\bibitem[Kim {\em et~al.}(2017)Kim, Madan, and Przytycka]{RN242}
Kim, Y.-A., Madan, S., and Przytycka, T.~M. (2017).
\newblock Wesme: uncovering mutual exclusivity of cancer drivers and beyond.
\newblock {\em Bioinformatics (Oxford, England)\/}, {\bf 33}(6), 814--821.

\bibitem[Kuleshov {\em et~al.}(2016)Kuleshov, Jones, Rouillard, {\em
  et~al.}]{RN245}
Kuleshov, M.~V., Jones, M.~R., Rouillard, A.~D., {\em et~al.} (2016).
\newblock Enrichr: a comprehensive gene set enrichment analysis web server 2016
  update.
\newblock {\em Nucleic Acids Res\/}, {\bf 44}(W1), W90--7.

\bibitem[Lamouille {\em et~al.}(2014)Lamouille, Xu, and Derynck]{XRN327}
Lamouille, S., Xu, J., and Derynck, R. (2014).
\newblock Molecular mechanisms of epithelial-mesenchymal transition.
\newblock {\em Nature reviews. Molecular cell biology\/}, {\bf 15}(3),
  178--196.

\bibitem[Lawrence {\em et~al.}(2013)Lawrence, Stojanov, Polak, {\em
  et~al.}]{RN233}
Lawrence, M.~S., Stojanov, P., Polak, P., {\em et~al.} (2013).
\newblock Mutational heterogeneity in cancer and the search for new
  cancer-associated genes.
\newblock {\em Nature\/}, {\bf 499}(7457), 214--218.

\bibitem[Leiserson {\em et~al.}(2015)Leiserson, Wu, Vandin, {\em
  et~al.}]{XRN235}
Leiserson, M. D.~M., Wu, H.-T., Vandin, F., {\em et~al.} (2015).
\newblock Comet: a statistical approach to identify combinations of mutually
  exclusive alterations in cancer.
\newblock {\em Genome Biology\/}, {\bf 16}(1), 160.

\bibitem[Linehan {\em et~al.}(2010)Linehan, Srinivasan, and Schmidt]{XRN247}
Linehan, W.~M., Srinivasan, R., and Schmidt, L.~S. (2010).
\newblock The genetic basis of kidney cancer: a metabolic disease.
\newblock {\em Nature Reviews Urology\/}, {\bf 7}, 277.

\bibitem[Mularoni {\em et~al.}(2016)Mularoni, Sabarinathan, Deu-Pons, {\em
  et~al.}]{XRN344}
Mularoni, L., Sabarinathan, R., Deu-Pons, J., {\em et~al.} (2016).
\newblock Oncodrivefml: a general framework to identify coding and non-coding
  regions with cancer driver mutations.
\newblock {\em Genome Biology\/}, {\bf 17}(1), 128.

\bibitem[Page {\em et~al.}(1998)Page, Brin, Motwani, {\em et~al.}]{XRN260}
Page, L., Brin, S., Motwani, R., {\em et~al.} (1998).
\newblock {\em The PageRank Citation Ranking: Bringing Order to the Web\/}.

\bibitem[Papaemmanuil {\em et~al.}(2016)Papaemmanuil, Gerstung, Bullinger, {\em
  et~al.}]{XRN230}
Papaemmanuil, E., Gerstung, M., Bullinger, L., {\em et~al.} (2016).
\newblock Genomic classification and prognosis in acute myeloid leukemia.
\newblock {\em New England Journal of Medicine\/}, {\bf 374}(23), 2209--2221.

\bibitem[Paull {\em et~al.}(2013)Paull, Carlin, Niepel, {\em et~al.}]{XRN254}
Paull, E.~O., Carlin, D.~E., Niepel, M., {\em et~al.} (2013).
\newblock Discovering causal pathways linking genomic events to transcriptional
  states using tied diffusion through interacting events (tiedie).
\newblock {\em Bioinformatics\/}, {\bf 29}(21), 2757--64.

\bibitem[Pham {\em et~al.}(2019)Pham, Liu, Bracken, {\em et~al.}]{RN228}
Pham, V. V.~H., Liu, L., Bracken, C.~P., {\em et~al.} (2019).
\newblock Cbna: A control theory based method for identifying coding and
  non-coding cancer drivers.
\newblock {\em PLOS Computational Biology\/}, {\bf 15}(12), e1007538.

\bibitem[Porta-Pardo {\em et~al.}(2015)Porta-Pardo, Hrabe, and Godzik]{XRN285}
Porta-Pardo, E., Hrabe, T., and Godzik, A. (2015).
\newblock Cancer3d: understanding cancer mutations through protein structures.
\newblock {\em Nucleic Acids Research\/}, {\bf 43}(Database issue), D968--D973.

\bibitem[Poulos {\em et~al.}(2015)Poulos, Sloane, Hesson, {\em et~al.}]{XRN294}
Poulos, R.~C., Sloane, M.~A., Hesson, L.~B., {\em et~al.} (2015).
\newblock The search for cis-regulatory driver mutations in cancer genomes.
\newblock {\em Oncotarget\/}, {\bf 6}(32), 32509--32525.

\bibitem[Puente {\em et~al.}(2015)Puente, Bea, Valdes-Mas, {\em
  et~al.}]{XRN297}
Puente, X.~S., Bea, S., Valdes-Mas, R., {\em et~al.} (2015).
\newblock Non-coding recurrent mutations in chronic lymphocytic leukaemia.
\newblock {\em Nature\/}, {\bf 526}, 519.

\bibitem[Rassenti {\em et~al.}(2017)Rassenti, Balatti, Ghia, {\em
  et~al.}]{XRN162}
Rassenti, L.~Z., Balatti, V., Ghia, E.~M., {\em et~al.} (2017).
\newblock Microrna dysregulation to identify therapeutic target combinations
  for chronic lymphocytic leukemia.
\newblock {\em Proceedings of the National Academy of Sciences\/}, {\bf
  114}(40), 10731.

\bibitem[Reimand and Bader(2013)Reimand and Bader]{RN234}
Reimand, J. and Bader, G.~D. (2013).
\newblock Systematic analysis of somatic mutations in phosphorylation signaling
  predicts novel cancer drivers.
\newblock {\em Molecular systems biology\/}, {\bf 9}, 637--637.

\bibitem[Rose {\em et~al.}(2013)Rose, Bi, Bluhm, {\em et~al.}]{XRN286}
Rose, P.~W., Bi, C., Bluhm, W.~F., {\em et~al.} (2013).
\newblock The rcsb protein data bank: new resources for research and education.
\newblock {\em Nucleic Acids Research\/}, {\bf 41}(Database issue), D475--D482.

\bibitem[Sakoparnig {\em et~al.}(2015)Sakoparnig, Fried, and
  Beerenwinkel]{XRN257}
Sakoparnig, T., Fried, P., and Beerenwinkel, N. (2015).
\newblock Identification of constrained cancer driver genes based on mutation
  timing.
\newblock {\em PLOS Computational Biology\/}, {\bf 11}(1), e1004027.

\bibitem[Stephens {\em et~al.}(2012)Stephens, Tarpey, Davies, {\em
  et~al.}]{XRN248}
Stephens, P.~J., Tarpey, P.~S., Davies, H., {\em et~al.} (2012).
\newblock The landscape of cancer genes and mutational processes in breast
  cancer.
\newblock {\em Nature\/}, {\bf 486}, 400.

\bibitem[Tamborero {\em et~al.}(2013)Tamborero, Gonzalez-Perez, and
  Lopez-Bigas]{XRN246}
Tamborero, D., Gonzalez-Perez, A., and Lopez-Bigas, N. (2013).
\newblock Oncodriveclust: exploiting the positional clustering of somatic
  mutations to identify cancer genes.
\newblock {\em Bioinformatics\/}, {\bf 29}(18), 2238--44.

\bibitem[Tokheim {\em et~al.}(2016)Tokheim, Papadopoulos, Kinzler, {\em
  et~al.}]{Tokheim14330}
Tokheim, C.~J., Papadopoulos, N., Kinzler, K.~W., {\em et~al.} (2016).
\newblock Evaluating the evaluation of cancer driver genes.
\newblock {\em Proceedings of the National Academy of Sciences\/}, {\bf
  113}(50), 14330--14335.

\bibitem[Vandin(2017)Vandin]{XRN233}
Vandin, F. (2017).
\newblock Computational methods for characterizing cancer mutational
  heterogeneity.
\newblock {\em Front Genet\/}, {\bf 8}, 83.

\bibitem[Vinayagam {\em et~al.}(2011)Vinayagam, Stelzl, Foulle, {\em
  et~al.}]{RN235}
Vinayagam, A., Stelzl, U., Foulle, R., {\em et~al.} (2011).
\newblock A directed protein interaction network for investigating
  intracellular signal transduction.
\newblock {\em Sci Signal\/}, {\bf 4}(189), rs8.

\bibitem[Vinayagam {\em et~al.}(2016)Vinayagam, Gibson, Lee, {\em
  et~al.}]{XRN165}
Vinayagam, A., Gibson, T.~E., Lee, H.-J., {\em et~al.} (2016).
\newblock Controllability analysis of the directed human protein interaction
  network identifies disease genes and drug targets.
\newblock {\em Proceedings of the National Academy of Sciences\/}, {\bf
  113}(18), 4976.

\bibitem[Wang {\em et~al.}(2014)Wang, Mezlini, Demir, {\em et~al.}]{RN251}
Wang, B., Mezlini, A.~M., Demir, F., {\em et~al.} (2014).
\newblock Similarity network fusion for aggregating data types on a genomic
  scale.
\newblock {\em Nature Methods\/}, {\bf 11}(3), 333--337.

\bibitem[Wang {\em et~al.}(2010)Wang, Lu, Qiu, {\em et~al.}]{RN240}
Wang, J., Lu, M., Qiu, C., {\em et~al.} (2010).
\newblock Transmir: a transcription factor-microrna regulation database.
\newblock {\em Nucleic Acids Res\/}, {\bf 38}(Database issue), D119--22.

\bibitem[Weinhold {\em et~al.}(2014)Weinhold, Jacobsen, Schultz, {\em
  et~al.}]{XRN298}
Weinhold, N., Jacobsen, A., Schultz, N., {\em et~al.} (2014).
\newblock Genome-wide analysis of non-coding regulatory mutations in cancer.
\newblock {\em Nature genetics\/}, {\bf 46}(11), 1160--1165.

\bibitem[Wendl {\em et~al.}(2011)Wendl, Wallis, Lin, {\em et~al.}]{XRN253}
Wendl, M.~C., Wallis, J.~W., Lin, L., {\em et~al.} (2011).
\newblock Pathscan: a tool for discerning mutational significance in groups of
  putative cancer genes.
\newblock {\em Bioinformatics\/}, {\bf 27}(12), 1595--602.

\bibitem[Wong {\em et~al.}(2018)Wong, Chen, Chen, {\em et~al.}]{XRN295}
Wong, N.~W., Chen, Y., Chen, S., {\em et~al.} (2018).
\newblock Oncomir: an online resource for exploring pan-cancer microrna
  dysregulation.
\newblock {\em Bioinformatics\/}, {\bf 34}(4), 713--715.

\bibitem[Xu {\em et~al.}(2017a)Xu, Wu, Cheng, {\em et~al.}]{HRN16}
Xu, L., Wu, W., Cheng, G., {\em et~al.} (2017a).
\newblock Enhancement of proliferation and invasion of gastric cancer cell by
  kdm5c via decrease in p53 expression.
\newblock {\em Technology in cancer research \& treatment\/}, {\bf 16}(2),
  141--149.

\bibitem[Xu {\em et~al.}(2017b)Xu, Le, Liu, {\em et~al.}]{RN250}
Xu, T., Le, T.~D., Liu, L., {\em et~al.} (2017b).
\newblock Cancersubtypes: an r/bioconductor package for molecular cancer
  subtype identification, validation and visualization.
\newblock {\em Bioinformatics\/}, {\bf 33}(19), 3131--3133.

\bibitem[Zhang {\em et~al.}(2011)Zhang, Baran, Cros, {\em et~al.}]{XRN276}
Zhang, J., Baran, J., Cros, A., {\em et~al.} (2011).
\newblock International cancer genome consortium data portal: a one-stop shop
  for cancer genomics data.
\newblock {\em Database\/}, {\bf 2011}, bar026--bar026.

\bibitem[Zhang {\em et~al.}(2013)Zhang, Zhang, Wang, {\em et~al.}]{XRN228}
Zhang, J., Zhang, S., Wang, Y., {\em et~al.} (2013).
\newblock Identification of mutated core cancer modules by integrating somatic
  mutation, copy number variation, and gene expression data.
\newblock {\em BMC Syst Biol\/}, {\bf 7 Suppl 2}, S4.

\bibitem[Zhang {\em et~al.}(2019)Zhang, Pham, Liu, {\em et~al.}]{RN230}
Zhang, J., Pham, V. V.~H., Liu, L., {\em et~al.} (2019).
\newblock Identifying mirna synergism using multiple-intervention causal
  inference.
\newblock {\em BMC Bioinformatics\/}, {\bf 20}(23), 613.

\end{thebibliography}

\newpage

\newpage

\end{document}